\documentclass[12pt]{article} 
\pdfoutput=1
\usepackage{graphicx,floatflt,amssymb,rotate}
\usepackage{mathrsfs}
\usepackage{amssymb}
\usepackage{amsfonts}
\usepackage{amsmath}
\usepackage{color}
\usepackage{graphicx}
\usepackage{subfig}
\usepackage{multirow}
\usepackage{float}
\usepackage{pstricks}
\usepackage[toc,page]{appendix}
\usepackage{cite}
\usepackage{soul}
\usepackage[colorlinks=true,
linkcolor=red,
urlcolor=blue,
citecolor=blue]{hyperref}
\usepackage{color}

\def\R1{\varepsilon_1}
\def\E8{\varepsilon_8}

\newcommand{\bea}{\begin{eqnarray}}
\newcommand{\eea}{\end{eqnarray}}
\newcommand{\bd}{\begin{displaymath}}
\newcommand{\ed}{\end{displaymath}}

\newcommand{\be}{\begin{equation}}
\newcommand{\ee}{\end{equation}}
\newcommand{\bi}{\begin{itemize}}
\newcommand{\ei}{\end{itemize}}

\begin{document}
\vskip 30pt  
 
\begin{center}  
{\Large \bf Belle II Constraints on the Non-Minimal Universal Extra Dimensional Model} \\
\vspace*{1cm}  
\renewcommand{\thefootnote}{\fnsymbol{footnote}}  
{{\sf Avirup Shaw$^1$\footnote{email: avirup.cu@gmail.com}} 
}\\  
\vspace{10pt}  
{{\em $^1$Department of Physics, Kalna College,\\ Affiliated to The Burdwan University,\\
Near old bus stand, Kalna, West Bengal 713409, India}}
\normalsize  
\end{center} 

\begin{abstract}
\noindent 
{Recent measurements by the Belle II Collaboration report a branching ratio for the decay $B^+\to K^+ \nu \bar{\nu}$ with a 3.5$\sigma$ significance and deviating by 2.7$\sigma$ from Standard Model expectation. In this article, we interpret this experimental result through the lens of a Non-minimal Universal Extra Dimensional model, where the mass profiles and interactions among various Kaluza-Klein states differ remarkably from those in the minimal scenario. Our analysis indicates that, according to the model's parameters, the lower limit on the inverse of the compactification radius has been elevated to approximately 900 GeV. In contrast, conducting the same analysis within the model variant in which all boundary terms are nullified does not permit the determination of a lower bound on the inverse compactification radius.
}
\vskip 5pt \noindent 
\end{abstract}

\renewcommand{\thesection}{\Roman{section}}  
\setcounter{footnote}{0}  
\renewcommand{\thefootnote}{\arabic{footnote}}

\section{Introduction}
The identification of the Higgs boson by the Large Hadron Collider (LHC) in 2012 represented a significant advancement for the Standard Model (SM) of particle physics \cite{Aad:2012tfa,Chatrchyan:2012xdj}. Furthermore, experimental data across diverse categories have consistently aligned with the SM predictions
for several decades. Nevertheless, there are still unresolved challenges, including the mystery
of dark matter (DM), the baryon asymmetry of the universe (BAU), and the origin of neutrino mass,
which the SM does not adequately address. Additionally, in flavour physics, experimental
findings for various observables reveal significant discrepancies when compared with SM
predictions. As a result, a fundamental objective of particle physics is to identify new
particles and interactions, collectively referred to as ``new physics" (NP), with the aim of developing
definitive solutions to the aforementioned challenges. There are two main strategies for
investigating NP. The first strategy, positioned at the high-energy frontier, involves the direct
production of these new degrees of freedom. Conversely, the second strategy, positioned at
the high-precision frontier, emphasises the examination of the indirect virtual effects
connected to these novel particles. In this latter approach, NP would endeavour to bridge the
disparity between the predictions of the SM and the results obtained from experimental
measurements.


Focusing on the second approach to identifying NP, observables related to Flavour Changing Neutral Current (FCNC) processes—defined by quark transitions that proceed without any alteration in electric charge—constitute highly sensitive indicators for physics beyond the SM (BSM). This sensitivity arises because FCNC processes are strongly suppressed within the SM framework due to the Glashow-Iliopoulos-Maiani (GIM) mechanism. Consequently, numerous FCNC-decay related observables have been measured by various experimental collaborations, including Belle, BABAR, and LHCb. In several cases, these experimental findings have revealed significant deviations from SM predictions. Such discrepancies have been explored within a variety of BSM theoretical frameworks, including supersymmetry, extra-dimensional models, extensions involving additional gauge symmetries, and scenarios incorporating discrete symmetries. 

Recently, the Belle II collaboration, utilising a dataset corresponding to an integrated luminosity of 362 fb\(^{-1}\), has reported the first evidence for the decay \(B^+ \to K^+ \nu \bar{\nu}\). The measured branching ratio, obtained by combining results from both inclusive and hadronic tagging methods, is given by~\cite{Belle-II:2023esi}:
\begin{align}\label{eq:exp:Belle II}
    {\text {Br}}(B^+ \to K^+ \nu \bar \nu)_{\text{Belle II}} & = \left[23 \pm 5(\text{stat})^{+5}_{-4}(\text{syst}) \right] \times 10^{-6}\;.
\end{align}
Notably, the measured value exceeds the SM prediction by approximately $2.7\sigma$. This intriguing result distinctly unveils a novel avenue for investigating NP, particularly noteworthy as it may suggest the presence of NP phenomena in the \( b \rightarrow s \nu \bar{\nu} \) transitions. Various theoretical frameworks have been proposed to account for the excess observed by the Belle II experiment in terms of NP effects \cite{Allwicher:2023xba,Bause:2023mfe,Altmannshofer:2023hkn,McKeen:2023uzo,Fridell:2023ssf,Gabrielli:2024wys,Felkl:2023ayn,Wang:2023trd,He:2023bnk,Datta:2023iln,Ho:2024cwk,Chen:2024jlj,Hou:2024vyw,He:2024iju,Bolton:2024egx,Buras:2024ewl,Altmannshofer:2024kxb,Marzocca:2024hua,Hu:2024mgf,Allwicher:2024ncl,Berezhnoy:2023rxx,Calibbi:2025rpx,Lee:2025jky,Athron:2023hmz,Rosauro-Alcaraz:2024mvx, Bolton:2025fsq, Aliev:2025hyp, Fuentes-Martin:2020hvc}. In particular, the decay rate of $B^+\rightarrow K^+\nu\bar{\nu}$ can be substantially modified in models incorporating particles beyond the SM. Accordingly, the present study interprets the branching ratio ${\text {Br}}(B^+\to K^+ \nu \bar{\nu})$ reported by the Belle II collaboration \cite{Belle-II:2023esi} within the context of a BSM framework known as the Non-Minimal Universal Extra Dimension (NMUED) model \cite{Dvali:2001gm, Carena:2002me, delAguila:2003bh, delAguila:2003kd, delAguila:2003gv, Schwinn:2004xa, Flacke:2008ne, Datta:2012xy, Flacke:2013pla}.

The Universal Extra Dimensions (UED) \cite{Appelquist:2000nn} model extends the SM by introducing a flat extra spatial dimension (denoted as $y$) accessible to all SM fields. The dimension $y$ is compactified on a circle ($S^1$) with a radius ($R$).  In this manifold, the fields are expanded in terms of four-dimensional (4D) Kaluza-Klein (KK) modes, where the zero-mode corresponds to the relevant SM field. The emergence of chiral fermions within the SM is facilitated by the imposition of a discrete symmetry \( Z_2 \) (where \( y \) is transformed to \( -y \)) on the extra spatial dimension. Consequently, this extra dimension is referred to as an \( S^1/Z_2 \) orbifold, leading to a physical domain that extends from \( y = 0 \) to \( y = \pi R \). The \( y \leftrightarrow -y \) symmetry manifests as a conserved parity, termed KK-parity, which is defined as \( (-1)^n \), where \( n \) represents the KK-number and quantifies the discretised momentum along the \( y \)-direction. KK-parity conservation guarantees that the lightest KK particle (LKP) remains stable and does not decay into SM particles. As a result, the LKP is considered a viable candidate for DM within this theoretical framework \cite{Servant:2002hb, Servant:2002aq, Cheng:2002ej, Majumdar:2002mw, Burnell:2005hm, Kong:2005hn, Kakizaki:2006dz, Belanger:2010yx}. Furthermore, various extensions of this model have the potential to address several shortcomings of the SM, including issues related to gauge coupling unification \cite{Dienes:1998vh, Dienes:1998vg, Bhattacharyya:2006ym}, neutrino mass \cite{Hsieh:2006qe, Fujimoto:2014fka}, and the hierarchy of fermion masses \cite{Huang:2012kz}.

At the $n^{\text {th}}$ KK-level, the mass of the KK-partner corresponding to any SM particle can be represented by the expression $\sqrt{(m^2+(nR^{-1})^2)}$, where \( m \) denotes the zero-mode mass (the mass of the SM particle) and is significantly smaller than \( R^{-1} \). Consequently, the UED framework exhibits a nearly degenerate mass spectrum at each KK-level. This characteristic leads to a deficiency in the phenomenological relevance of the UED scenario, particularly in collider experiments. However, radiative corrections can address the issue of mass spectrum degeneracy. These corrections can be categorised into two types: bulk corrections, which are finite and non-zero solely for KK-excitations of gauge bosons, and boundary localised corrections, which are proportional to terms dependent on a logarithmically defined cut-off scale (\( \Lambda \)) \cite{Cheng:2002iz}. Boundary correction terms can be incorporated as 4D mass, kinetic, and other interaction terms for the KK-excited states at the two fixed boundary points (i.e., \( y = 0 \) and \( y = \pi R \)) of the orbifold. The inclusion of such terms is particularly natural within an extra-dimensional framework like UED, as they serve as counterterms for loop-induced contributions that depend on the cut-off scale $\Lambda$. In the minimal UED (MUED) model, a specific assumption is made whereby the boundary terms are selected to ensure that the 5D radiative corrections disappear at the cut-off scale\footnote{UED is treated as an effective field theory, defined by the presence of a cut-off scale $\Lambda$.} \( \Lambda \). Although this assumption is unique, it can be relaxed, allowing for the parametrisation of radiative corrections as kinetic, mass, and other interaction terms localised at the two fixed boundary points. This alternative formulation is referred to as the NMUED model. Within this context, in addition to the compactification radius (\( R \)), the coefficients of various boundary localised terms (BLTs) are treated as free parameters, which can be constrained by a range of experimental data pertaining to various physical observables. The literature contains numerous phenomenological studies within this framework. For instance, constraints on the coefficients of the BLTs have been derived from analyses of electroweak observables \cite{Flacke:2008ne, Flacke:2013pla}, the S, T, and U parameters \cite{delAguila:2003gv, Flacke:2013nta}, DM relic density \cite{Bonnevier:2011km, Datta:2013nua}, the production and decay of the SM Higgs boson \cite{Dey:2013cqa}, and collider studies from LHC experiments \cite{Datta:2012tv, Datta:2013yaa, Datta:2013lja, Shaw:2014gba, Shaw:2017whr, Ganguly:2018pzs}. Additional studies have focused on processes such as \( Z \to b\bar{b} \) \cite{Jha:2014faa}, branching ratios of rare $B$-meson decay processes (e.g., \( B_s \to \mu^+ \mu^- \) \cite{Datta:2015aka}, \( B \to X_s \gamma \) \cite{Datta:2016flx}, and \( B \to X_s \ell^+ \ell^- \)) \cite{Shaw:2019fin}, $\mathcal{R}_{D^{(*)}}$ anomalies \cite{Biswas:2017vhc, Dasgupta:2018nzt, Lee:2019phf}, flavour-changing rare top decays \cite{Dey:2016cve, Chiang:2018oyd}, $\Delta B=2$ transitions \cite{Shaw:2020fxf} and the unitarity of scattering amplitudes involving KK-excitations \cite{Jha:2016sre}.


This study presents the first comprehensive analysis of the branching ratio \( \mathrm{Br}(B^+ \to K^+ \nu \bar{\nu}) \) within the framework of the NMUED model. We analyse the contributions from KK-modes to the relevant Feynman diagrams, including on one-loop box diagrams. The functions derived from these diagrams are affected not only by the compactification radius but also by the parameters related to the BLT coefficients. Utilising these functions, we will compute the relevant Wilson Coefficient (WC), which will facilitate the estimation of the corresponding branching ratios within the NMUED framework. By comparing our theoretical predictions with the currently allowed ranges of the branching ratios, we aim to effectively constrain the parameter space of the NMUED model. Additionally, our analysis enables us to establish a lower limit on \( R^{-1} \) and to compare this result with findings from our previous research \cite{Datta:2015aka, Datta:2016flx, Shaw:2019fin}. 

The organisation of this paper is outlined as follows. Section \ref{model} will offer a brief overview of the NMUED model. Following this, section \ref{btonunu} will discuss the computational considerations related to the $B^+\to K^+\nu\bar{\nu}$ transition within the NMUED framework. Section \ref{anls} will focus on the presentation of our numerical results. Lastly, section \ref{concl} will provide a summary of the findings.

\section{KK-parity conserving NMUED framework}\label{model}
This section provides the key characteristics of the NMUED model, specifically those relevant for our current investigation. Comprehensive treatments of this framework can be found in the literature \cite{Dvali:2001gm, Carena:2002me, delAguila:2003bh, delAguila:2003kd, delAguila:2003gv, Schwinn:2004xa, Flacke:2008ne, Datta:2012xy, Datta:2012tv, Datta:2013yaa, Datta:2013lja, Shaw:2014gba, Shaw:2017whr, Ganguly:2018pzs, Jha:2014faa, Datta:2015aka, Datta:2016flx, Shaw:2019fin, Biswas:2017vhc, Shaw:2020fxf}. KK-parity conservation is maintained by imposing symmetry on the boundary terms at $y=0$ and $y=\pi R$. This symmetry leads to the presence of a stable LKP, making it a viable DM candidate—most often the first KK-excitation of the photon. In-depth studies on DM phenomenology within this scenario are available in \cite{Datta:2013nua}.

We illustrate a 5D fermion action that includes boundary localised kinetic terms (BLKTs), characterised by the coefficient $r_f$ \cite{Schwinn:2004xa, Datta:2013nua, Datta:2015aka, Datta:2016flx, Shaw:2019fin, Biswas:2017vhc, Shaw:2020fxf}:
\begin{align}
S_{\text{fermion}} = \int d^5x \Big[ &\bar{\Psi}_L i\Gamma^M D_M \Psi_L + r_f \{\delta(y) + \delta(y - \pi R)\} \bar{\Psi}_L i\gamma^\mu D_\mu P_L \Psi_L \nonumber \\
&+ \bar{\Psi}_R i\Gamma^M D_M \Psi_R + r_f \{\delta(y) + \delta(y - \pi R)\} \bar{\Psi}_R i\gamma^\mu D_\mu P_R \Psi_R \Big].
\label{factn}
\end{align}
\newpage
In this context, \(\Psi_L(x, y)\) and \(\Psi_R(x, y)\) represent four-component 5D Dirac fermions, without distinction between quark or lepton fields. It is important to emphasise that the subscripts \(L\) and \(R\) do not indicate chirality; rather, they are employed solely as a notational convention. Each of these spinors is further decomposed into two-component spinors according to the following scheme: \cite{Schwinn:2004xa, Datta:2013nua, Datta:2015aka, Datta:2016flx, Shaw:2019fin, Biswas:2017vhc, Shaw:2020fxf}:
\begin{equation} 
\Psi_L(x,y) = \begin{pmatrix}\phi_L(x,y) \\ \chi_L(x,y)\end{pmatrix}
=   \sum_n \begin{pmatrix}\phi^{(n)}_L(x) f_L^n(y) \\ \chi^{(n)}_L(x) g_L^n(y)\end{pmatrix}, 
\label{fermionexpnsn1}
\end{equation}
\begin{equation} 
\Psi_R(x,y) = \begin{pmatrix}\phi_R(x,y) \\ \chi_R(x,y) \end{pmatrix} 
=   \sum_n \begin{pmatrix}\phi^{(n)}_R(x) f_R^n(y) \\ \chi^{(n)}_R(x) g_R^n(y) \end{pmatrix}. 
\label{fermionexpnsn2} 
\end{equation}

The corresponding wavefunctions for KK-modes take the following form \cite{Carena:2002me, Flacke:2008ne, Datta:2013nua, Datta:2015aka, Datta:2016flx, Shaw:2019fin, Biswas:2017vhc, Shaw:2020fxf}:
\begin{eqnarray}
f_L^n = g_R^n = N^f_n \left\{ \begin{array}{rl}
                \displaystyle \frac{\cos\left[m_{f^{(n)}} \left (y - \frac{\pi R}{2}\right)\right]}{\cos[ \frac{m_{f^{(n)}} \pi R}{2}]}  &\mbox{for $n$ even,}\\
                \displaystyle \frac{{-}\sin\left[m_{f^{(n)}} \left (y - \frac{\pi R}{2}\right)\right]}{\sin[ \frac{m_{f^{(n)}} \pi R}{2}]} &\mbox{for $n$ odd,}
                \end{array} \right.
                \label{flgr}
\end{eqnarray}
and
\begin{eqnarray}
g_L^n =-f_R^n = N^f_n \left\{ \begin{array}{rl}
                \displaystyle \frac{\sin\left[m_{f^{(n)}} \left (y - \frac{\pi R}{2}\right)\right]}{\cos[ \frac{m_{f^{(n)}} \pi R}{2}]}  &\mbox{for $n$ even,}\\
                \displaystyle \frac{\cos\left[m_{f^{(n)}} \left (y - \frac{\pi R}{2}\right)\right]}{\sin[ \frac{m_{f^{(n)}} \pi R}{2}]} &\mbox{for $n$ odd.}
                \end{array} \right.
\end{eqnarray}

The KK-mode wavefunctions are normalised using the modified inner product, accounting for boundary terms as follows \cite{Carena:2002me, Flacke:2008ne, Datta:2013nua, Datta:2015aka, Datta:2016flx, Shaw:2019fin, Biswas:2017vhc, Shaw:2020fxf}:
\begin{equation}\label{orthonorm}
\begin{aligned}
&\left.\begin{array}{r}
                  \int_0 ^{\pi R}
dy \; \left[1 + r_{f}\{ \delta(y) + \delta(y - \pi R)\}\right]f_L^mf_L^n\\
                  \int_0 ^{\pi R}
dy \; \left[1 + r_{f}\{ \delta(y) + \delta(y - \pi R)\}\right]g_R^mg_R^n
\end{array}\right\}=&&\delta^{n m}~;
&&\left.\begin{array}{l}
                 \int_0 ^{\pi R}
dy \; f_R^mf_R^n\\
                 \int_0 ^{\pi R}
dy \; g_L^mg_L^n
\end{array}\right\}=&&\delta^{n m}~,
\end{aligned}
\end{equation}

yielding the normalisation constant \cite{Carena:2002me, Flacke:2008ne, Datta:2013nua, Datta:2015aka, Datta:2016flx, Shaw:2019fin, Biswas:2017vhc, Shaw:2020fxf}:
\begin{equation}\label{norm}
N^f_n=\sqrt{\frac{2}{\pi R}}\Bigg[ \frac{1}{\sqrt{1 + \frac{r^2_f m^2_{f^{(n)}}}{4} + \frac{r_f}{\pi R}}}\Bigg].
\end{equation}

The KK-mass $m_{f^{(n)}}$ is determined from \cite{Carena:2002me, Datta:2013nua, Datta:2015aka, Datta:2016flx, Shaw:2019fin, Biswas:2017vhc, Shaw:2020fxf}:
\begin{eqnarray}
  \frac{r_{f} m_{f^{(n)}}}{2}= \left\{ \begin{array}{rl}
         -\tan \left(\frac{m_{f^{(n)}}\pi R}{2}\right) &\mbox{for $n$ even,}\\
          \cot \left(\frac{m_{f^{(n)}}\pi R}{2}\right) &\mbox{for $n$ odd.}
          \end{array} \right.   
          \label{fermion_mass}      
\end{eqnarray}

In this context, we aim to examine the Yukawa interactions, as the heavy mass of the top quark is instrumental in amplifying the quantum effects addressed in the current study. The Yukawa interaction in 5D, including boundary localised contributions characterised by $r_y$, takes the form \cite{Datta:2015aka, Datta:2016flx, Shaw:2019fin, Biswas:2017vhc, Shaw:2020fxf}:
\begin{equation}
S_{\text{Yukawa}} = -\int d^5x \left[ \lambda_5^t \bar{\Psi}_L \tilde{\Phi} \Psi_R + r_y\{\delta(y) + \delta(y - \pi R)\} \lambda_5^t \bar{\phi}_L \tilde{\Phi} \chi_R + \text{h.c.} \right]\;,
\label{yukawa}
\end{equation}
where $\tilde{\Phi} = i \sigma_2 \Phi^*$ and $\Phi=\left(\begin{array}{cc} \phi^+\\\phi^0\end{array}\right)$ is the 5D Higgs doublet. $\lambda^5_t$ denotes the 5D coupling strength associated with the Yukawa interaction for the third generation.

We can expand the actions (given in Eqs.~\ref{factn} and \ref{yukawa}) using Eqs.~\ref{fermionexpnsn1} and \ref{fermionexpnsn2}, which consequently yield a mass matrix at the KK-level $n$ for the third-generation quarks \cite{Datta:2015aka, Datta:2016flx, Shaw:2019fin, Biswas:2017vhc, Shaw:2020fxf}:

\begin{equation}
\label{fermion_mix}
-\begin{pmatrix}
\bar{\phi_L}^{(n)} & \bar{\phi_R}^{(n)}
\end{pmatrix}
\begin{pmatrix}
m_{f^{(n)}}\delta^{nm} & m_{t} {\mathscr{I}}^{nm}_1 \\ m_{t} {\mathscr{I}}^{mn}_2& -m_{f^{(n)}}\delta^{mn}
\end{pmatrix}
\begin{pmatrix}
\chi^{(m)}_L \\ \chi^{(m)}_R
\end{pmatrix}+{\rm h.c.}\;.
\end{equation}
Here, \( m_t \) denotes the mass of the SM top quark, while $m_{f^{(n)}}$ is obtained from the solution of the transcendental equations presented in Eq.\;\ref{fermion_mass}. The overlap integrals are defined by \cite{Datta:2015aka, Datta:2016flx, Shaw:2019fin, Biswas:2017vhc, Shaw:2020fxf}:
  \[ {\mathscr{I}}^{nm}_1=\left(\frac{1+\frac{r_f}{\pi R}}{1+\frac{r_y}{\pi R}}\right)\times\int_0 ^{\pi R}\;dy\;
\left[ 1+ r_y \{\delta(y) + \delta(y - \pi R)\} \right] g_{R}^m f_{L}^n,\] \;\;{\rm and}\;\;\[{\mathscr{I}}^{nm}_2=\left(\frac{1+\frac{r_f}{\pi R}}{1+\frac{r_y}{\pi R}}\right)\times\int_0 ^{\pi R}\;dy\;
 g_{L}^m f_{R}^n .\]
 
The integral ${\mathscr{I}}^{nm}_1$ is non-zero for both cases where \( n = m \) and \( n \neq m \). Nonetheless, in the limit as \( r_y \) approaches \( r_f \), this integral yields a value of one when \( n = m \) and zero when \( n \neq m \). Conversely, the integral ${\mathscr{I}}^{nm}_2$ is non-zero exclusively for the case where \( n = m \) and also equals one when \( r_y = r_f \). It is important to note that, in our analysis, we have opted for the condition of equality \( (r_y = r_f) \) in order to avoid the complexities associated with mode mixing and to facilitate the construction of a more straightforward fermion mixing matrix \cite{Jha:2014faa, Datta:2015aka, Datta:2016flx, Shaw:2019fin, Biswas:2017vhc}. 
This criterion\footnote{Nevertheless, it is generally permissible to utilise unequal coefficients for the boundary terms associated with kinetic and Yukawa interactions in the context of fermions.} has been uniformly implemented across the entirety of our analysis.

Under the aforementioned equality condition ($r_y = r_f$), the mass matrix, as presented in Eq.\;\ref{fermion_mix}, can be readily diagonalised through the application of the following bi-unitary transformations for both the left-handed and right-handed fields \cite{Datta:2015aka, Datta:2016flx, Shaw:2019fin, Biswas:2017vhc, Shaw:2020fxf}:
\begin{equation}
U_{L}^{(n)}=\begin{pmatrix}
\cos\alpha_{tn} & \sin\alpha_{tn} \\ -\sin\alpha_{tn} & \cos\alpha_{tn}
\end{pmatrix},~~U_{R}^{(n)}=\begin{pmatrix}
\cos\alpha_{tn} & \sin\alpha_{tn} \\ \sin\alpha_{tn} & -\cos\alpha_{tn}
\end{pmatrix}.
\end{equation}
Here, $\alpha_{tn}\left[ = \frac12\tan^{-1}\left(\frac{m_{t}}{m_{f^{(n)}}}\right)\right]$ is defined as the mixing angle. The gauge eigen states $\Psi_L(x,y)$ and $\Psi_R(x,y)$ can be expressed clearly in terms of mass eigen states $T^1_t$ and $T^2_t$ through the subsequent relationships \cite{Datta:2015aka, Datta:2016flx, Shaw:2019fin, Biswas:2017vhc, Shaw:2020fxf}:

\vspace*{-1cm}
\begin{equation}
\begin{tabular}{p{8cm}p{8cm}}
{\begin{align}
&{\phi^{(n)}_L} =  \cos\alpha_{tn}T^{1(n)}_{tL}-\sin\alpha_{tn}T^{2(n)}_{tL},\nonumber \\
&{\chi^{(n)}_L} =  \cos\alpha_{tn}T^{1(n)}_{tR}+\sin\alpha_{tn}T^{2(n)}_{tR},\nonumber
\end{align}}
&
{\begin{align}
&{\phi^{(n)}_R} =  \sin\alpha_{tn}T^{1(n)}_{tL}+\cos\alpha_{tn}T^{2(n)}_{tL},\nonumber \\
&{\chi^{(n)}_R} =  \sin\alpha_{tn}T^{1(n)}_{tR}-\cos\alpha_{tn}T^{2(n)}_{tR}. \nonumber
\end{align}}
\end{tabular}
\end{equation}
The mass eigen states \( T^{1(n)}_t \) and \( T^{2(n)}_t \) exhibit the same mass eigen value at each KK-level. Note that, for the \( n^{{\text {th}}} \) KK-level, the mass eigen value is expressed as \( M_{t^{(n)}} \equiv \sqrt{m_{t}^{2} + m_{f^{(n)}}^{2}} \).

The gauge and scalar sectors are now considered. The kinetic terms of the gauge and scalar fields receive boundary corrections with respective coefficients $r_W$, $r_B$, and $r_\phi$. These can be incorporated into the 5D actions as \cite{Flacke:2008ne, Datta:2014sha, Jha:2014faa, Datta:2015aka, Datta:2016flx, Shaw:2019fin, Biswas:2017vhc, Dey:2016cve, Shaw:2020fxf}:
\begin{align}
S_{\text{gauge}} = -\frac{1}{4} \int d^5x &\Big[ W^a_{MN} W^{aMN} + r_W \{\delta(y) + \delta(y - \pi R)\} W^a_{\mu\nu} W^{a\mu\nu} \nonumber \\
&+ B_{MN} B^{MN} + r_B \{\delta(y) + \delta(y - \pi R)\} B_{\mu\nu} B^{\mu\nu} \Big]\;,
\label{pure-gauge}
\end{align}
and 
\begin{equation}
S_{\text{scalar}} = \int d^5x \Big[ (D_M \Phi)^\dagger (D^M \Phi) + r_\phi \{\delta(y) + \delta(y - \pi R)\} (D_\mu \Phi)^\dagger (D^\mu \Phi) \Big]\;.
\label{higgs}
\end{equation}

The 5D field strength tensors are defined as:
\begin{align}
W^a_{MN} &= \partial_M W^a_N - \partial_N W^a_M - \tilde{g}_2 \epsilon^{abc} W^b_M W^c_N\;, \\
B_{MN} &= \partial_M B_N - \partial_N B_M\;.
\label{ugfs}
\end{align}

The gauge fields $W^a_M (\equiv W^a_\mu, W^a_4)$ and $B_M (\equiv B_\mu, B_4)$ ($M=0,1 \ldots 4$) correspond to the $SU(2)_L$ and $U(1)_Y$ gauge symmetries, respectively. The 5D covariant derivative takes the form:
\begin{equation}
D_M = \partial_M + i \tilde{g}_2 \frac{\sigma^a}{2} W^a_M + i \tilde{g}_1 \frac{Y}{2} B_M\;,
\end{equation}
with 5D gauge coupling constants ${\tilde{g}_2}$ and ${\tilde{g}_1}$. The generators of the gauge groups \( SU(2)_L \) and \( U(1)_Y \) are expressed as \( \frac{\sigma^{a}}{2} \) (where \( a \equiv 1, 2, 3 \)) and \( \frac{Y}{2} \), respectively. Each of the gauge and scalar fields associated with the aforementioned actions (Eqs.\;\ref{pure-gauge} and \ref{higgs}) can be represented through suitable KK-wavefunctions as follows \cite{Datta:2014sha, Jha:2014faa, Datta:2015aka, Datta:2016flx, Shaw:2019fin, Biswas:2017vhc, Dey:2016cve, Shaw:2020fxf}:
\begin{equation}\label{Amu}
V_{\mu}(x,y)=\sum_n V_{\mu}^{(n)}(x) a^n(y),\;\;\;\;\
V_{4}(x,y)=\sum_n V_{4}^{(n)}(x) b^n(y)\;,
\end{equation}
\vspace*{-0.5cm}
and
\begin{equation}\label{chi}
\Phi(x,y)=\sum_n \Phi^{(n)}(x) h^n(y),
\end{equation}
where, both the 5D $SU(2)_L$ and $U(1)_Y$ gauge bosons are typically represented by the notation $(V_\mu, V_4)$.

For the neutral gauge bosons, additional complexity arises due to mixing between $B$ and $W^3$ components in both the bulk and boundary sectors. Diagonalisation of the corresponding kinetic terms is only tractable if one assumes\footnote{Nevertheless, it is generally permissible to consider the scenario where \( r_W \neq r_B \). In this context, the mixing between the \( B \) and \( W^3 \) fields, both in the bulk and at the boundary points, results in the emergence of off-diagonal elements within the neutral gauge boson mass matrix.} $r_W = r_B$ \cite{Datta:2014sha, Jha:2014faa, Datta:2015aka, Datta:2016flx, Shaw:2019fin, Biswas:2017vhc, Dey:2016cve}, which simplifies the neutral sector similarly to the MUED case. Under this condition, the first KK-level mixing of $B^{(1)}$ and $W^{3(1)}$ yields $\gamma^{(1)}$ and $Z^{(1)}$. The photon’s first KK-excitation, $\gamma^{(1)}$, is stable due to KK-parity and acts as the LKP, making it an excellent DM candidate \cite{Datta:2013nua}.

To proceed consistently with quantisation, the gauge fixing terms must also include the boundary terms. This leads to a gauge-fixing action of the form \cite{Datta:2014sha, Jha:2014faa, Datta:2015aka, Datta:2016flx, Shaw:2019fin, Biswas:2017vhc, Dey:2016cve, Shaw:2020fxf}:
\begin{align}
S_{\text{gf}} = -\frac{1}{\xi_y} \int d^5x &\left|\partial^\mu W^+_\mu + \xi_y \left( \partial_y W^+_4 + i M_W \phi^+ [1 + r_V (\delta(y) + \delta(y - \pi R))] \right) \right|^2 \nonumber \\
& - \frac{1}{2\xi_y} \int d^5x \left[ \partial^\mu Z_\mu + \xi_y (\partial_y Z_4 - M_Z \chi [1 + r_V (\delta(y) + \delta(y - \pi R))]) \right]^2 \nonumber \\
& - \frac{1}{2\xi_y} \int d^5x [\partial^\mu A_\mu + \xi_y \partial_y A_4]^2\;.
\label{gauge-fix}
\end{align}

Here, $M_W$ and $M_Z$ are the usual SM masses of $W^\pm$ and $Z$ bosons. For a comprehensive examination of the gauge fixing action and mechanism within the framework of NMUED, we direct the reader to the work \cite{Datta:2014sha}. The action presented in Eq.\;\ref{gauge-fix} is both complex and crucial for the NMUED context, particularly as we aim to compute one-loop diagrams, which are essential for the current analysis, utilising the Feynman gauge. Due to the BLKTs, the extra dimensional profiles of the fields acquire a non-uniform weight, necessitating adjustments in gauge fixing. This non-homogeneity requires the introduction of a $y$-dependent gauge fixing parameter, denoted as $\xi_y$, as depicted in previous studies \cite{Datta:2014sha, Jha:2014faa, Datta:2015aka, Datta:2016flx, Shaw:2019fin, Biswas:2017vhc, Dey:2016cve, Shaw:2020fxf}:
\begin{equation}\label{gf_para}
\xi =\xi_y\,(1+ r_V\{ \delta(y) +  \delta(y - \pi R)\})\;,
\end{equation}
where $\xi$ remains independent of $y$. This relationship can be interpreted as a form of renormalisation of the gauge fixing parameter, as the BLKTs effectively serve as counterterms, accommodating the unknown ultraviolet corrections that arise in loop calculations. This redefinition (given in Eq.\;\ref{gf_para}) allows one to interpret $\xi_y$ as the bare gauge fixing parameter and $\xi$ as the renormalised parameter, taking values like $0$ (Landau gauge), $1$ (Feynman gauge), or $\infty$ (Unitary gauge).

The condition $r_V = r_\phi$ is essential to ensure the gauge fixing procedure works consistently in the presence of BLKTs. Under this requirement, the KK-masses of the gauge and scalar fields are identical (i.e., $m_{V^{(n)}}(=m_{\phi^{(n)}})$), and the mass of the charged gauge bosons at the $n^{{\text {th}}}$ KK-level is \cite{Datta:2014sha, Jha:2014faa, Datta:2015aka, Datta:2016flx, Shaw:2019fin, Biswas:2017vhc, Dey:2016cve, Shaw:2020fxf}:
\begin{equation}
M_{W^{(n)}} = \sqrt{M_{W}^{2}+m^2_{V^{(n)}}}\;.
\end{equation}

Likewise, in the ’t-Hooft Feynman gauge, the Goldstone bosons $G^{\pm(n)}$ that correspond to $W_\mu^{\pm(n)}$ possess the same mass $M_{W^{(n)}}$ \cite{Datta:2014sha, Jha:2014faa, Datta:2015aka, Datta:2016flx, Shaw:2019fin, Biswas:2017vhc, Dey:2016cve, Shaw:2020fxf}.

The KK-decomposition of interaction terms involves overlap integrals that arise from integrating the $y$-dependent wavefunctions. These integrals modify the effective couplings between zero-mode SM fields and KK-states, thereby distinguishing NMUED predictions from those of MUED. Explicit forms for these overlap integrals are provided in Appendix\;\ref{appA}. Further discussions are available in \cite{Datta:2014sha, Jha:2014faa, Datta:2015aka, Datta:2016flx, Shaw:2019fin, Biswas:2017vhc, Dey:2016cve, Shaw:2020fxf}.

\section{\boldmath$B\to K^{(\ast)} \nu \bar{\nu}$ transition in the NMUED scenario}\label{btonunu}
The decay process \( B^{+} \to K^+ \nu \bar{\nu} \) can be described at the quark level by the transition \( b \to s \nu \bar{\nu} \). The effective Hamiltonian\footnote{Note that in the present study, the NP scale is significantly higher than the $B$-meson mass scale.} that regulates this transition, as discussed in the literature \cite{Colangelo:1996ay, Altmannshofer:2009ma, Buras:2014fpa}, can be appropriately adjusted within the framework of the NMUED model. Hence, utilising the descriptions proposed in \cite{Buras:2002ej}, the effective Hamiltonian can be written as follows:
\begin{eqnarray}\label{hdb2}
{\cal H}^{b\to s\nu\bar{\nu}}_{\rm eff}=-\frac{4G_F}{\sqrt{2}}V_{tb}V^\ast_{ts}\Bigg(-\frac{X(x_t, r_f,r_V, R^{-1})}{s^2_W}\Bigg)\frac{e^2}{16\pi^2}[\bar s\gamma_\mu P_L b][\bar \nu\gamma^\mu (1-\gamma_5)\nu] + {\rm h. c.}\;. 
\end{eqnarray}
In the above equation $G_F$ is designated as the Fermi constant. The product $V_{tb}V^\ast_{ts}$ is the Cabibbo Kobayashi Maskawa (CKM) entry appropriate for the transition $b\to s\nu\bar{\nu}$ . The factor (-$\frac {X} {s^2_W}$) plays the role of WC that incorporates next-to-leading-order Quantum Chromo Dynamics (QCD) corrections \cite{Buchalla:1993bv, Buchalla:1998ba, Misiak:1999yg} and two-loop electroweak corrections \cite{Brod:2010hi}. The value of $s^2_W(=\sin^2{\theta_W})$ is 0.231 \cite{ParticleDataGroup:2024cfk}, where $\theta_W$ is the Weinberg angle.
\begin{figure}[H]
	\begin{center}
		\includegraphics[scale=0.6,angle=0]{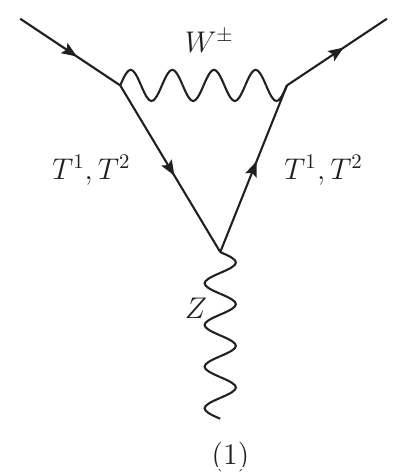}
		\includegraphics[scale=0.6,angle=0]{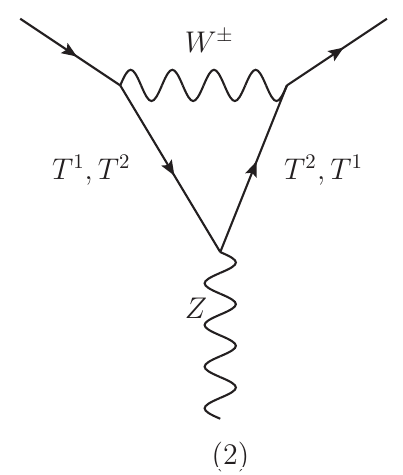}
		\includegraphics[scale=0.6,angle=0]{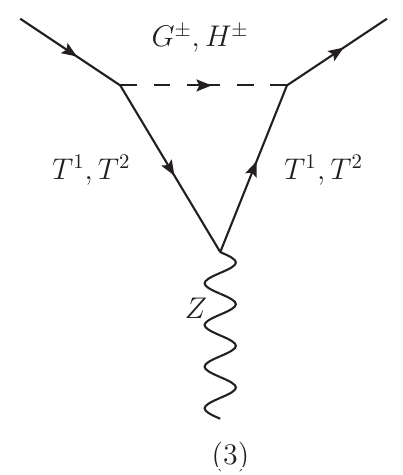}
		\includegraphics[scale=0.6,angle=0]{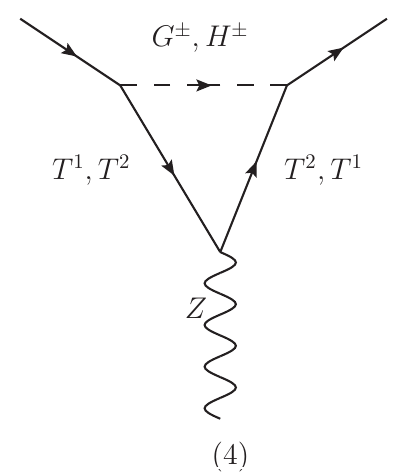}
		\includegraphics[scale=0.6,angle=0]{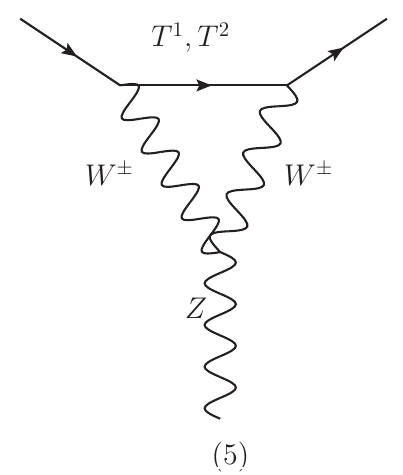}
		\includegraphics[scale=0.6,angle=0]{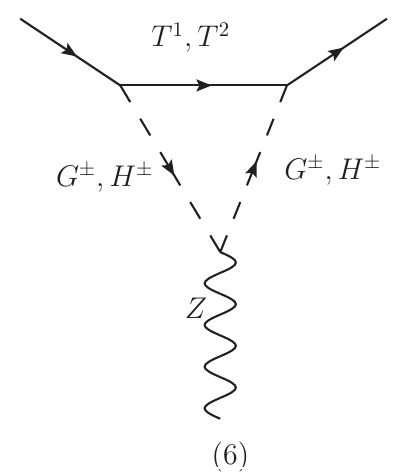}
		\includegraphics[scale=0.6,angle=0]{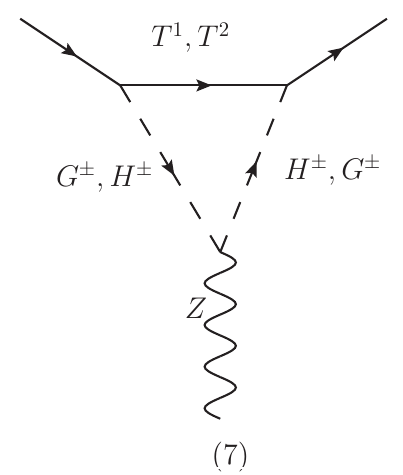}
		\includegraphics[scale=0.6,angle=0]{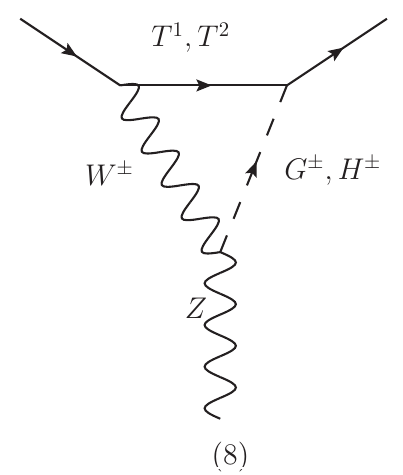}
		\caption{The decay process \( B^+ \to K^+ \nu \bar{\nu} \) receives the contributions from the \( Z \)-penguin diagrams.}
		\label{pen}
	\end{center}
\end{figure}

\begin{figure}[H]
	\begin{center}
		\includegraphics[scale=0.6,angle=0]{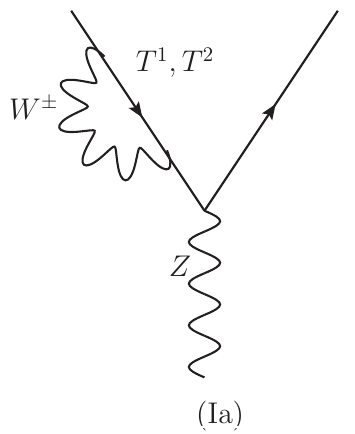}
		\includegraphics[scale=0.6,angle=0]{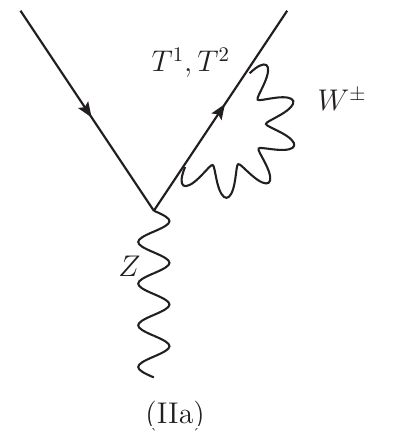}
		\includegraphics[scale=0.6,angle=0]{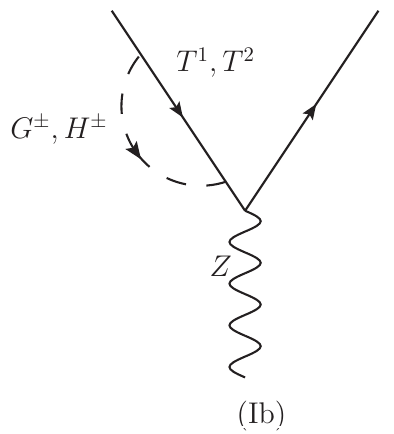}
		\includegraphics[scale=0.6,angle=0]{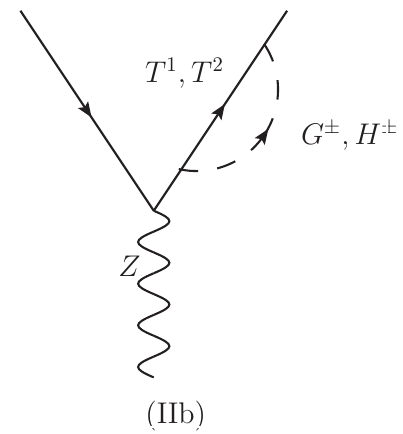}
		\caption{The self-energy diagrams contribute to the decay process \( B^+ \to K^+ \nu \bar{\nu} \).}
		\label{self}
	\end{center}
\end{figure}

For the relevant transition ($B^+ \to K^+ \nu \bar{\nu}$), the overall contribution—comprising both SM and KK-mode effects—in the NMUED framework is described by the function $X(x_t, r_f, r_V, R^{-1})$, which can be decomposed into two functions as shown below:
%
\be\label{Xfunction}
X(x_t,r_f,r_V,R^{-1})=C(x_t,r_f,r_V,R^{-1})+B^{\nu\bar{\nu}}(x_t,r_f,r_V,R^{-1})\;.
\ee
In the above expression, the function \( C(x_t, r_f, r_V, R^{-1}) \) arises from penguin (Fig.\;\ref{pen}) and self-energy (Fig.\;\ref{self}) diagrams and can be expressed as:
\begin{equation}\label{cn}
	C(x_t, r_f, r_V, R^{-1})=C_0(x_t)+
	\sum_{n=1}^\infty C_n(x_{t(n)},x_{u(n)})\;,
\end{equation}
where, $C_0(x_t)$ represents the corresponding SM contribution \cite{Buchalla:1995vs, Buras:2002ej}:
\begin{equation}
	C_0(x_t)={\frac{x_t}{8}}\left[\frac{x_t-6}{x_t-1}+\frac{3x_t+2}
	{(x_t-1)^2}\;\ln x_t\right]\;.
	\label{c0-eq}
\end{equation}
The pertinent contributions to the KK-mode arising from the NMUED framework are represented by the second term, which is derived from the $Z$-penguin diagrams and self-energy diagrams, as illustrated in Figs.\;\ref{pen} and \ref{self}. The function \( C_n(x_{t(n)}, x_{u(n)}) \) can be expressed as follows \cite{Buras:2002ej}: 
\begin{equation}
	C_n(x_{t(n)},x_{u(n)})= F(x_{t(n)})-F(x_{u(n)})\;.
	\label{subtract}
\end{equation}
The functions $F(x_{t(n)})$ and $F(x_{u(n)})$ are derived from the contributions of the modes $T^{1(n)}_{t}$, $T^{2(n)}_{t}$ and $T^{1(n)}_{u}$, $T^{2(n)}_{u}$ respectively. The function $F(x_{t(n)})$ is expressed in the following manner \cite{Buras:2002ej}:
\begin{equation}
	F(x_{t(n)}) = \sum_{i=1}^8 F_i(x_{t(n)}) + \left(\frac12 - \frac13
	\sin^2 \theta_W \right) \sum_{i=1}^2 \Delta S_i ( x_{t(n)})\;.
\end{equation}
Furthermore, \( F_i \) represents the contributions arising from the \( Z \)-penguin diagrams (1-8 in Fig. \ref{pen}), while \( \Delta S_i \) denotes the contributions from self-energy diagrams (Ia-IIb in Fig. \ref{self}), which are essential for the computation of the electroweak counterterms. The function \( F(x_{u(n)}) \) is defined as follows:
\begin{equation}
	F(x_{u(n)})=F(x_{t(n)})\bigg|_{x_{t}\rightarrow 0}.
\end{equation}
In the aforementioned equation (Eq.\;\ref{subtract}), the terms $F(x_{u(n)})$ are subtracted to account for the contributions arising from the KK-excitations of the first two generations of quarks in the $Z$-penguin and self-energy diagrams, utilising the GIM mechanism. In the context of the NMUED framework, the expression for $F(x_{t(n)})$ (given in Eq.\;\ref{fn}) is significantly distinct from that in the UED model, primarily due to the inclusion of the overlap integrals $I^n_1$ and $I^n_2$ \cite{Datta:2015aka}. The corresponding expressions can be found in Appendix\;\ref{appA}, specifically in Eqs.~\ref{i1} and \ref{i2}.

Within the NMUED framework, the function \( F(x_{t(n)}) \) is represented by the following expression:
{\small
	\begin{eqnarray}\label{fn}
		\hspace*{-2cm}
		&& F(x_{t(n)})=\left[ \frac{1}{8}\left\{-\ln{M^2_{t^{(n)}}}-\frac{3}{2}+h_q\left(x_{t(n)}\right)\right\}-\frac{1}{4} x_{t(n)} h_q\left(x_{t(n)}\right)\frac{m^2_{f^{(n)}}}{M^2_{t^{(n)}}}\right](I^n_1)^2
		\nonumber \\ &&
		-\frac{3}{4}\left[-\ln{M^2_{W^{(n)}}}-\frac{1}{6}-x_{t(n)}h_w\left(x_{t(n)}\right)\right](I^n_1)^2
		-\frac{1}{2}h_w\left(x_{t(n)}\right)\frac{m^2_{f^{(n)}}}{M^2_{W^{(n)}}}(I^n_1)^2
		\nonumber \\ &&
		+\bigg[\frac{1}{16}\left\{-\ln{M^2_{t^{(n)}}}-\frac{1}{2}+h_q\left(x_{t(n)}\right)\right\}(I^n_2)^2
		-\frac{1}{8}x_{t(n)}h_q\left(x_{t(n)}\right)\left\{\frac{m^4_{f^{(n)}}}{M^2_{t^{(n)}}m^2_{V^{(n)}}}+\frac{m^4_t}{M^2_{t^{(n)}}M^2_W}\right\}(I^n_1)^2\bigg]
		\nonumber \\ &&
		-\frac{1}{16}\left[2(I^n_2)^2+x_t(I^n_1)^2\right]\left[-\ln{M^2_{W^{(n)}}}+\frac{1}{2}-x_{t(n)}h_w\left(x_{t(n)}\right)\right]
		\nonumber \\ &&
		+\frac18 \left[
		-\frac{1}{2} \left\{\frac{1+x_{t(n)}}{1-x_{t(n)}} +  \frac{2
			x_{t(n)}^2 \ln x_{t(n)} }{(1-x_{t(n)})^2} \right\} - 
		\ln {M^2_{W^{(n)}}} \right](I^n_1)^2
		\nonumber \\ && 
		+\frac{1}{16} \left[
		(I^n_2)^2 + x_t(I^n_1)^2 \right]  \left[\frac{1}{2} 
		\left\{ \frac{1-3x_{t(n)}}{1-x_{t(n)}} -
		\frac{2 x_{t(n)}^2\ln x_{t(n)} }{(1-x_{t(n)})^2} \right\}   - 
		\ln {M^2_{W^{(n)}}} \right].
	\end{eqnarray}
}
The formulations for \( h_q \) and \( h_w \) can be found in Appendix \ref{appA} (Eqs. \ref{hq} and \ref{hw}).

\begin{figure}[H]
	\begin{center}
		\includegraphics[scale=0.8,angle=0]{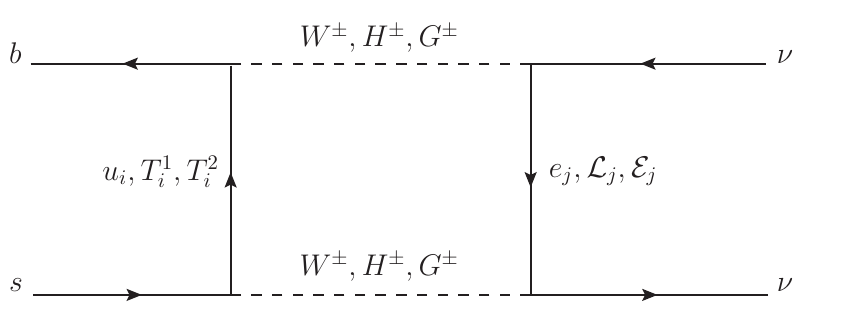}
		\caption{Box diagrams that play a role in the decay process $B^+\to K^+\nu\bar{\nu}$. The indices "$i$" and "$j$" represent fermion flavour indices.}
		\label{box}
	\end{center}
\end{figure}

The function \( B^{\nu\bar{\nu}}(x_t, r_f, r_V, R^{-1}) \), presented in Eq.\;\ref{Xfunction}, is defined by\footnote{We would like to point out that a similar expression for the UED model is available in \cite{Buras:2002ej}.}:
\begin{equation}\label{bn}
	B^{\nu\bar{\nu}}(x_t, r_f, r_V, R^{-1})=-4B_0(x_t)+
	\sum_{n=1}^\infty B^{\nu\bar{\nu}}_n(x_{t(n)}, x_{u(n)}, x_{e(n)})\;.
\end{equation}
The function \( B_0(x_t) \) represents the box diagram contribution within the SM \cite{Buchalla:1995vs, Buras:2002ej}:
\begin{equation}
	B_0(x_t)=\frac{1}{4}\left[\frac{x_t}{1-x_t}+\frac{x_t}
	{(x_t-1)^2}\;\ln x_t\right]\;.
	\label{b0-eq}
\end{equation}
The second term represents the overall contribution from the KK-modes, derived from the box diagrams shown in Fig.~\ref{box}. The function \( B^{\nu\bar{\nu}}_n(x_{t(n)}, x_{u(n)}, x_{\nu(n)}) \) is defined as follows:
\begin{equation}\label{b_n}
	B^{\nu\bar{\nu}}_n(x_{t(n)}, x_{u(n)}, x_{e(n)})= G(x_{t(n)},x_{e(n)})-G(x_{u(n)},x_{e(n)})\;.
\end{equation}
Within the NMUED framework, the expression for \( G(x_{t(n)}, x_{e(n)}) \) is written as\footnote{To the best of our knowledge, the current article presents, for the first time, the expression derived from the box diagrams (shown in Fig.~\ref{box}) relevant to the decay process $B^+ \to K^+ \nu \bar{\nu}$ in the context of the NMUED model.}:
{\small
	\begin{eqnarray}\label{gn}
		\hspace*{-2cm}
		&&G(x_{t(n)},x_{e(n)})=(I^n_1)^4\Bigg[-\frac{M_W^2}{M^2_{W^{(n)}}}~ U(x_{t(n)}, x_{e(n)})
		+\frac12 \frac{M_W^2 M^2_{t^{(n)}}m^2_{f^{(n)}}}{M^6_{W^{(n)}}}~ \widetilde{U}(x_{t(n)}, x_{e(n)})
		\nonumber \\ &&
		+\frac12 \frac{M_W^2 m^2_{f^{(n)}}}{M^6_{W^{(n)}}}\left(\frac{M^2_W m^2_{f^{(n)}}}{m^2_{V^{(n)}}}-m^2_t\right)~ \widetilde{U}(x_{t(n)}, x_{e(n)})
		-\frac18 \frac{M_W^2 m^2_{f^{(n)}}}{M^6_{W^{(n)}}}\left(\frac{M^2_W m^2_{f^{(n)}}}{m^2_{V^{(n)}}}-m^2_t\right)~ U(x_{t(n)}, x_{e(n)})
		\nonumber \\ &&
		-\frac{1}{16} \frac{M_W^2 M^2_{t^{(n)}}m^2_{f^{(n)}}}{M^6_{W^{(n)}}}~U(x_{t(n)}, x_{e(n)})
		-\frac{1}{16} \frac{M_W^2}{M^6_{W^{(n)}}}\left(\frac{M^4_Wm^4_{f^{(n)}}}{m^4_{V^{(n)}}}+m^2_tm^2_{f^{(n)}}\right)~U(x_{t(n)}, x_{e(n)})\Bigg],
	\end{eqnarray}
}
and the function \( G(x_{u(n)}, x_{e(n)}) \) is defined as follows:
\begin{equation}
	G(x_{u(n)},x_{e(n)})=G(x_{t(n)},x_{e(n)})\bigg|_{x_{t}\rightarrow 0}.
\end{equation}

The formulations for the functions \( U \) and \( \widetilde{U} \) are presented in Appendix \ref{appA} (Eqs. \ref{U1} and \ref{U3}).

As shown in Eq.~\ref{b_n}, the expression for \( B^{\nu\bar{\nu}}_n \) incorporates contributions from KK-excitations of the first two quark generations into the analytical framework.

In all the previous expressions, \( x_t = \frac{m_t^2}{M_W^2} \), \( x_{V{(n)}} = \frac{m_{V^{(n)}}^2}{M_W^2} \), and \( x_{f{(n)}} = \frac{m_{f^{(n)}}^2}{M_W^2} \), where \( m_{V^{(n)}} \) and \( m_{f^{(n)}} \) can be derived from the solution of the transcendental equation given in Eq.~\ref{fermion_mass}. 

To this end, we are now equipped to evaluate the branching fraction of the decay process \( B^+ \to K^+ \nu \bar{\nu} \) in the NMUED framework. Differential branching fraction for the decay process \( B^+ \to K^+ \nu \bar{\nu} \) can be expressed as follows \cite{Colangelo:1996ay, Altmannshofer:2009ma, Buras:2014fpa}:
\begin{equation}\label{brBtoKnunubar}
\frac{\mathrm{d}{\text {Br}}}{\mathrm{d}q^2} (B^+ \to K^+ \nu \bar{\nu}) = \tau_{B^{+}} \frac{G_F^2 \alpha_{\mathrm{em}}^2}{256 \pi^5} \frac{\lambda_K^{3/2}}{m_B^3}\frac{X^2}{S^4_W}|V_{tb}V^\ast_{ts}|^2 [f_+(q^2)]^2,
\end{equation}
where $\alpha_{\mathrm{em}}(= \frac{e^2}{4\pi})$ is the fine structure constant and $\lambda_K= m_B^4  + m_{K}^4 + q^4 - 2 (m_B^2 m_{K}^2+ m_{K}^2 q^2  + m_B^2 q^2)$. \( 0 < q^2 \leq (m_B - m_K)^2 \) represents the invariant mass of the di-neutrinos. Moreover, \( f_+(q^2) \) denotes the vector form factor for the \( B^+ \to K^+ \) transition, which has been discussed in Appendix~\ref{BFF} for convenience. The total branching fraction can be obtained by integrating the differential branching fraction over the entire kinematically allowed region.

It is important to highlight a significant aspect of this article. While the primary objective is to analyse the recent Belle II data \cite{Belle-II:2023esi} within the framework of the NMUED, the effective Hamiltonian presented in Eq.\;\ref{hdb2} for the \( b \to s \nu\bar{\nu} \) transition also governs several other important observables. As for example the branching fraction of the decay process \( B \to K^{\ast} \nu \bar{\nu} \) (denoted as \( \text{Br}(B \to K^{\ast} \nu \bar{\nu}) \)). Furthermore, it imposes a constraint derived from the upper limit of the branching fraction reported by the Belle collaboration \cite{Belle:2017oht}. Consequently, the following discussion will focus on the evaluation of the observable.

Analogous to the previous case, the branching fraction for the decay process \( (B \to K^{\ast} \nu \bar{\nu}) \) can be expressed as \cite{Colangelo:1996ay, Altmannshofer:2009ma, Buras:2014fpa}:
\begin{equation}\label{brBtoKstnunubar}
\frac{\mathrm{d}{\text {Br}}}{\mathrm{d}q^2} (B \to K^{\ast} \nu \bar{\nu}) = \tau_{B^0} \frac{G_F^2 \alpha_{\mathrm{em}}^2}{256 \pi^5} \frac{\lambda_{K^\ast}^{1/2}}{m_B^3}\frac{X^2}{S^4_W}|V_{tb}V^\ast_{ts}|^2 \mathbb{F}(q^2),
\end{equation}
with \( \lambda_{K^\ast}=m_B^4  + m_{K^\ast}^4 + q^4 - 2 (m_B^2 m_{K^\ast}^2+ m_{K^\ast}^2 q^2  + m_B^2 q^2) \), and the function \( \mathbb{F}(q^2) \) is defined as:
\begin{equation}\label{Ffunc}
\mathbb{F}(q^2) = 2q^2(m_B + m_{K^\ast})^2[A_1(q^2)]^2 + 64 m_{K^\ast}^2 m_B^2 [A_{12}(q^2)]^2 + \frac{2q^2\lambda_{K^\ast}}{(m_B + m_{K^\ast})^2} [V(q^2)]^2.
\end{equation}
For the sake of convenience, the form factors \( A_1(q^2) \), \( A_{12}(q^2) \), and \( V(q^2) \), associated with the \( B \to K^\ast \) transition, are presented in Appendix~\ref{BFF}. The overall branching fraction can be determined by integrating the differential branching fraction across the entire kinematically permissible region.

Before we proceed to the numerical analysis, we would like to emphasise some important issues. It has been previously noted that the presence of various BLTs in the NMUED action results in nontrivial modifications to the KK-masses and couplings (altered by \( I^n_1 \) and \( I^n_2 \)) associated with KK-excitations, in contrast to their UED counterparts\footnote{It is important to highlight that, concerning the NMUED model, the quantity of Feynman diagrams corresponds exactly to that calculated in the UED model, as elaborated in reference \cite{Buras:2002ej}.}. Therefore, it is not feasible to derive the function \( X_n \) ( defined as \( X_n = C_n + B^{\nu\bar{\nu}}_n \)) in the NMUED context merely by rescaling the corresponding values from the UED model \cite{Buras:2002ej}. Consequently, we have independently calculated the KK-mode contributions within the NMUED framework utilising Feynman diagrams (illustrated in Figs. \ref{pen}, \ref{self}, and \ref{box}). Furthermore, it is evident from Eqs. \ref{fn} and \ref{gn} that the KK-mode contributions in the NMUED model differ significantly from those in the UED formulation. However, if we set the boundary terms to zero (i.e., \( r_V = 0 \) and \( r_f = 0 \)), we can readily recover the results of the UED model from our expressions. Moreover, as the KK-number increases (i.e., for larger values of \( n \)), the masses of the fields \( T^{1(n)}_t \), \( T^{2(n)}_t \), \( T^{1(n)}_u \), and \( T^{2(n)}_u \) become nearly degenerate. This mass degeneracy leads to a suppression of the function \( X_n \), for higher values of \( n \). Therefore, under the NMUED framework, only a limited number of terms in the KK-mode summation have a substantial impact. Additionally, it is important to note that in the one-loop Feynman diagram calculations, we only consider interactions where the zero-mode field couples to pairs of KK-excitations with identical KK-numbers. In the context of KK-parity conserving NMUED, it is also possible to have non-zero interactions involving KK-excitations with KK-numbers \( n \), \( m \), and \( p \), provided that \( n + m + p \) is an even integer. Nevertheless, we have explicitly confirmed that the final results remain largely unchanged even when accounting for all potential off-diagonal interactions \cite{Jha:2014faa, Datta:2015aka, Datta:2016flx, Shaw:2019fin, Shaw:2020fxf}.

\section{Numerical analysis}\label{anls}
In the present study, we have, for the first time, calculated the KK-contributions to the WC of the $B^+\to K^+\nu\bar{\nu}$ (and also $B\to K^\ast\nu\bar{\nu}$) transition within the framework of the NMUED scenario. The function $X(x_t,r_f,r_V,R^{-1})$, as defined in Eq.\;\ref{Xfunction}, encapsulates the dependence on the KK-masses of both gauge bosons and fermions in the NMUED context. Furthermore, in light of the analysis regarding the impact of the SM Higgs mass on vacuum stability within the UED model \cite{Datta:2012db}, we have considered the sum of KK-contributions up to 10 KK-levels\footnote{Electroweak vacuum stability \cite{Datta:2012db} constrains the cutoff scale in UED-like models to $\Lambda R=\mathcal{O}(5\!-\!10)$, as KK contributions drive the Higgs quartic coupling negative at higher scales. This implies that only a few low-lying KK-modes contribute, making truncated KK-sums theoretically consistent for phenomenological analyses.}. This approach contrasts with previous studies, which typically included contributions from 20 to 30 KK-levels. Subsequently, we combined the total KK-contributions with those from the SM\footnote{For the mass of the SM $W^\pm$ boson, we utilised $M_W=80.369$ GeV \cite{ParticleDataGroup:2024cfk}, and for the mass of the SM top quark, we adopted $\overline{m}_t(m_t)=172.56$ GeV \cite{ParticleDataGroup:2024cfk}.}. Additionally, due to the convergent nature of the KK-summation in UED type models with a single extra spatial dimension, particularly in the context of one-loop calculations \cite{Dey:2004gb}, the numerical results are unlikely to vary significantly based on the inclusion of a greater number of KK-levels when evaluating the KK-contributions for the loop diagrams \cite{Datta:2015aka, Datta:2016flx, Shaw:2019fin, Shaw:2020fxf}. Moreover, it is important to note that the function \( X(x_t, r_f, r_V, R^{-1}) \) essentially serves as the WC for the specific \( b \to s \nu \bar{\nu} \) transition. Consequently, following the calculation of this function within the framework of the NMUED model, we have evaluated the numerical values of the observables associated with the \( b \to s \nu \bar{\nu} \) transitions by employing the package {\tt flavio} \cite{Straub:2018kue}. This package serves exclusively as a numerical tool for the evaluation of flavour observables, depending upon the provision of the pertinent low-energy WC. All short-distance effects specific to the NMUED model are analytically incorporated into modified loop functions that contribute to the \( b \to s \) effective Hamiltonian. These functions are independently calculated within the NMUED model and subsequently supplied as external inputs to \texttt{flavio} \cite{Straub:2018kue}.

\subsection{Constraints and permissible range for BLT parameter values}\label{const}
In this analysis, we emphasise the constraints that have been taken into account.
\begin{itemize}
\item 
In the context of the NMUED model, extensive analyses have been conducted on various rare $B$-decay processes characterised by FCNC, particularly focusing on the $b\to s$ transition. Notable decay processes include $B_s\rightarrow \mu^+\mu^-$ \cite{Datta:2015aka}, $B\rightarrow X_s\gamma$ \cite{Datta:2016flx}, and $B\rightarrow X_s\ell^+\ell^-$ \cite{Shaw:2019fin}. These decay modes have been instrumental in investigating potential NP scenarios. It is significant to note that the expressions for these observables rely on a common set of parameters, namely $r_V$, $r_f$, and $R^{-1}$, which are also relevant to the observables analysed in the current study. Consequently, we incorporate the branching ratios of these rare decay processes as constraints in our analysis. Furthermore, in this article we have derived the theoretical predictions for the branching ratios ${\rm Br}(B_s\rightarrow \mu^+\mu^-)$, ${\rm Br}(B\rightarrow X_s\gamma)$, and ${\rm Br}(B\rightarrow X_{_s}\ell^+\ell^-)$ as outlined in \cite{Datta:2015aka}, \cite{Datta:2016flx}, and \cite{Shaw:2019fin}, utilising the package {\tt flavio} \cite{Straub:2018kue}. The corresponding relevant updated experimental data have been obtained from the references \cite{ParticleDataGroup:2024cfk}, \cite{HFLAV:2022esi, ParticleDataGroup:2024cfk}, and \cite{Lees:2013nxa}, respectively.
\item Electroweak precision tests (EWPT) play a crucial role in constraining various forms of BSM physics. By employing the method of correcting the Fermi constant \( G_F \) at the tree level, one can derive corrections to the Peskin-Takeuchi parameters S, T, and U within the framework of the NMUED model. This characteristic distinguishes the NMUED scenario from the minimal version of the UED model, where such corrections arise at one-loop processes. A comprehensive analysis of EWPT within the NMUED model has been presented in the references \cite{Datta:2015aka, Biswas:2017vhc}. In the NMUED model, the S, T, and U parameters are expressed as functions of \( r_V \), \( r_f \), and \( R^{-1} \). Following the methodology outlined in the aforementioned references \cite{Datta:2015aka, Biswas:2017vhc}, we have incorporated EWPT as a constraint in our analysis.  

\end{itemize}

In this context, we aim to examine the range of values for the BLT parameters utilised in our current analysis. Generally, the BLT parameters can assume both negative and positive values. Nonetheless, as demonstrated by Eq.\;\ref{norm}, when the ratio ${r_f}/{R}$ is assigned the value \(-\pi\), the zero-mode solution becomes divergent. Furthermore, for values of ${r_f}/{R}$ lower than \(-\pi\), the zero-mode fields display properties analogous to those of ghost fields. Consequently, values of the BLT parameters that fall below $- \pi R$ should be excluded from consideration. Nevertheless, for the sake of thoroughness, we have incorporated some negative values of the BLT parameters in our numerical analysis. However, based on the study of electroweak precision data~\cite{Datta:2015aka, Biswas:2017vhc}, a large portion of negative values of the BLT parameters has been disfavoured.

In the numerical analysis, all theoretical and experimental constraints outlined above are applied concurrently. During the parameter space exploration, any parameter point that fails to meet at least one of these constraints is discarded from further analysis. Only those parameter points that comply with the entire set of constraints are retained and subsequently employed to compute the relevant flavour observables and to produce the numerical results presented in the figures, such as Fig.~\ref{blkt_bran}. This methodological approach is consistent with the procedure utilised in our previous investigations of $B$-physics observables within the NMUED framework, as detailed in the references \cite{Datta:2015aka, Datta:2016flx, Shaw:2019fin, Shaw:2020fxf}. 
\subsection{Results}
{At this stage, considering the previously outlined constraints, our objective is to depict the parameter space that complies with the 99\% confidence level (CL) bounds for the allowed range of the branching ratio of the decay process \( B^+ \to K^+ \nu \bar{\nu} \), while simultaneously respecting the upper limit imposed on the branching ratio of the decay \( B \to K^\ast \nu \bar{\nu} \). In the current version of the NMUED framework, we have three independent free parameters: the inverse radius of compactification \( R^{-1} \), and the dimensionless scaled BLT parameters for bosons \( R_V (= r_V/R) \) and fermions \( R_f (= r_f/R) \). This can be elucidated as follows: as the values of the BLT parameters increase, the KK-masses decrease, which in turn enhances the loop function \( X \) derived from the one-loop Feynman diagrams depicted in Figs. \ref{pen}, \ref{self}, and \ref{box}. Consequently, to counterbalance this enhancement, it is necessary to increase the values of \( R^{-1} \), since the KK-masses increase with larger values of \( R^{-1} \). 
\begin{figure}[ht!]
\begin{center}
\includegraphics[scale=0.95,angle=0]{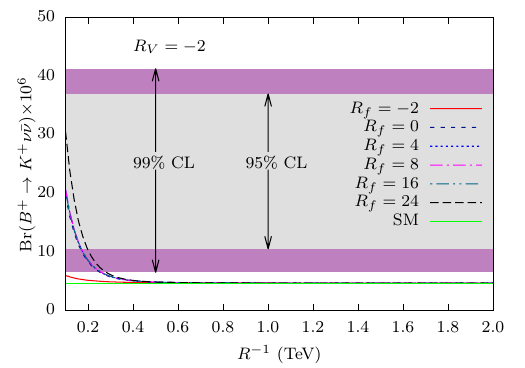}
\includegraphics[scale=0.95,angle=0]{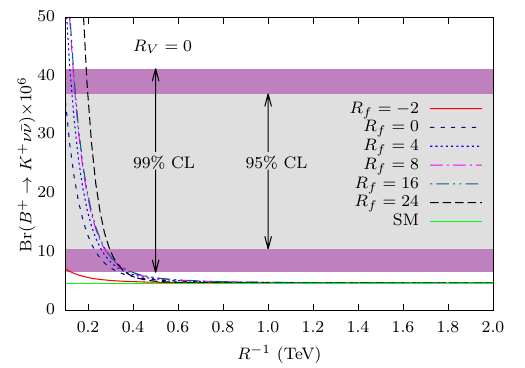}
\includegraphics[scale=0.95,angle=0]{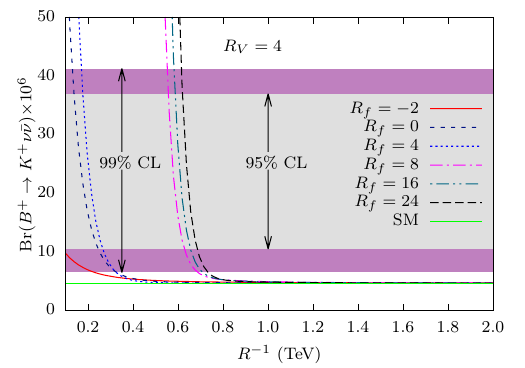}
\includegraphics[scale=0.95,angle=0]{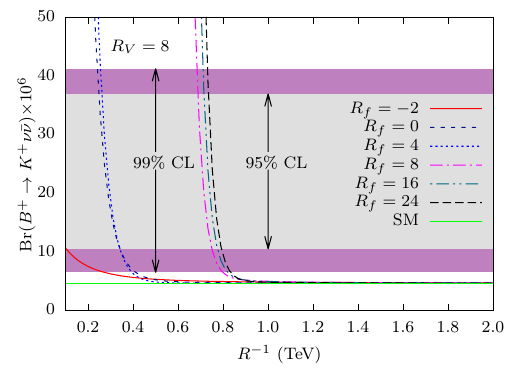}
\includegraphics[scale=0.95,angle=0]{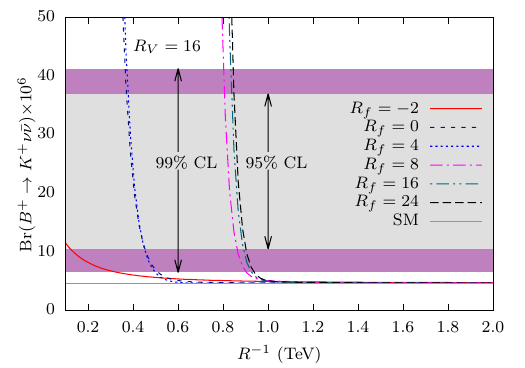}
\includegraphics[scale=0.95,angle=0]{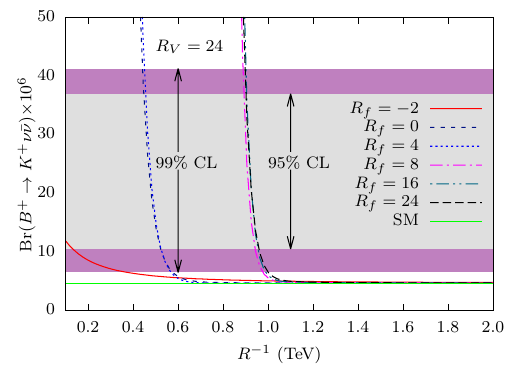}
\caption{The branching ratio of the decay process $B^+ \to K^+ \nu \bar{\nu}$ is analysed as a function of $R^{-1}$ (in TeV) for various values of $R_f(=r_f/R)$. The six panels correspond to distinct values of $R_V(=r_V/R)$. The figure exclusively presents parameter points that fulfill all the constraints outlined in subsection~\ref{const}. In the computation of the WCs, contributions from up to 10 KK-levels are incorporated through different loop functions. The horizontal grey (purple) band illustrates the 95\% (99\%) confidence interval of the experimentally determined branching ratio.}
\label{blkt_bran}
\end{center}
\end{figure}

Another significant aspect of the NMUED scenario is that, as a result of EWPT, the region where \( R_V \approx R_f \) and higher values of \( R^{-1} \) are generally disfavoured for most negative values of the BLT parameters \cite{Datta:2015aka, Biswas:2017vhc}. In the subsequent sections, for illustrative purposes, we have selected several sets of non-zero scaled BLT parameters \( (R_V, R_f) \) that are permissible under the different constraints considered in our analysis. 

The Fig.\;\ref{blkt_bran} comprises six panels that illustrate the relationship between the branching ratio ${\text {Br}}(B^+\to K^+\nu\bar{\nu})$ and the parameter \(R^{-1}\) across various combinations of BLT parameters, including cases where these parameters are set to zero. When a specific curve intersects the 99\% CL of the permissible range for the experimental value, we can derive the lower limits for \(R^{-1}\). These lower limit values are presented in Table \ref{lower_R} in units of GeV. The data in this table indicate that for multiple selections of BLT parameters, we have determined lower limits for \(R^{-1}\). For each specific BLT configuration, a lower limit on \( R^{-1} \) is established only when the corresponding theoretical curve crosses the upper experimental constraint. For instance, with \(R_V=8\) and \(R_f=16\), our analysis yields a lower limit of \(R^{-1} \approx 711\) GeV. In contrast, employing a different set of BLT parameters, specifically \(R_V=24\) and \(R_f=24\), results in a lower limit of \(R^{-1} \approx 902\) GeV. At the limit of KK-mode decoupling, all observable values reach saturation and exhibit no dependence on \(R^{-1}\). 

\begin{table}[htb!]
\begin{center}
\hspace*{-0.5cm}
\resizebox{17.5cm}{!}{
\begin{tabular}{||c||c|c||c|c||c|c||c|c||}
\hline
{} & \multicolumn{2}{|c||}{$R_V=4$} & \multicolumn{2}{|c||}{$R_V=8$} & \multicolumn{2}{|c||}{$R_V=16$} & \multicolumn{2}{|c||}{$R_V=24$}\\
\hline
{$R_f$} &
95\% CL & 99\% CL &
95\% CL & 99\% CL &
95\% CL & 99\% CL &
95\% CL & 99\% CL \\
\hline
0  & 137 & 131 & 248 & 242 & 371 & 365 & 447 & 441 \\
4  & 176 & 171 & 262 & 255 & 378 & 374 & 457 & 451 \\
8  & 557 & 552 & 688 & 684 & 807 & 803 & 892 & 888 \\
16 & 588 & 583 & 715 & 711 & 840 & 834 & 903 & 898 \\
24 & 619 & 615 & 736 & 731 & 848 & 845 & 905 & 902 \\
\hline
\end{tabular}
}
\end{center}
\caption{Lower bounds on the inverse compactification radius, \( R^{-1} \) (in GeV), have been derived from the branching ratio of the decay process \( B^+ \to K^+ \nu \bar{\nu} \). These constraints are presented for various values of the BLT parameters, corresponding to the $95\%$ and $99\%$ CL intervals of the experimental measurements reported by the Belle II collaboration~\cite{Belle-II:2023esi}. For the $99\%$ CL interval, the analysis additionally incorporates the constraint arising from the upper limit on the branching ratio of the decay \( B \to K^\ast \nu \bar{\nu} \), as reported by the Belle collaboration~\cite{Belle:2017oht}. The bounds on \( R^{-1} \) are extracted based on the experimental branching ratio specified in Eq.~1, and consequently, they carry an overall uncertainty on the order of $30\%$, predominantly stemming from the experimental measurement uncertainties.}
\label{lower_R}
\end{table}

Furthermore, it is important to clarify the impact of the $\mathrm{Br}(B \to K^\ast \nu \bar{\nu})$ constraint on the derived lower limits of $R^{-1}$, particularly in the context of the bounds presented in Table~\ref{lower_R}. In general, removing a constraint enlarges the allowed parameter space, although the resulting bounds on specific parameters may either strengthen or weaken depending on the interplay among different observables. In analyses involving multiple observables, each observable can probe distinct regions of the parameter space with varying sensitivities. In the present study, the experimental upper limit on $\mathrm{Br}(B \to K^\ast \nu \bar{\nu})$, as reported by the Belle Collaboration~\cite{Belle:2017oht}, favours regions corresponding to relatively smaller values of $R^{-1}$. Consequently, this constraint is comparatively less restrictive in the region preferred by the remaining observables, and its inclusion extends the combined allowed parameter space towards lower values of $R^{-1}$. When this constraint is removed, the allowed region is instead determined by the remaining observables, which tend to prefer comparatively larger values of $R^{-1}$.
As a result, although the total allowed parameter space is formally enlarged, the region favoured by the combined analysis shifts, leading to a marginal increase in the effective lower bound on $R^{-1}$. This behaviour reflects the non-trivial interplay between different observables, rather than a simple monotonic dependence on the number of constraints. For the specified sets of BLT parameters, the resulting lower bounds on $R^{-1}$ are approximately $715~\mathrm{GeV}$ and $905~\mathrm{GeV}$, respectively, both of which remain consistent with the $95\%$ confidence level allowed range for the branching ratio of $B^+ \to K^+ \nu \bar{\nu}$.

{

For the purpose of comparison, it is pertinent to note that in the UED scenario, electroweak precision measurements and direct collider searches \cite{Deutschmann:2017bth} generally establish a lower limit on the compactification scale of approximately \( R^{-1} \sim \mathcal{O}(1\,\text{TeV}) \). However, these limits are known to be influenced by the specific mass spectrum and decay patterns involved \cite{Deutschmann:2017bth}. In contrast, within the NMUED framework, the inclusion of BLTs can significantly alter the KK-mass spectrum and interaction strengths. This modification permits a relaxation of the aforementioned bounds, potentially lowering the $R^{-1}$ to a few hundred GeV, depending upon suitable selections of BLT parameters. Within this context, our analysis reveals that the theoretical predictions for the branching ratio of the decay \( B^+ \to K^+ \nu \bar{\nu} \) remain consistent with current experimental constraints across the entire parameter space that satisfies all imposed conditions. Therefore, although this decay channel offers valuable complementary insights, it does not, in the present analysis, independently yield the most stringent lower bound on \( R^{-1} \) when compared to the results presented in \cite{Shaw:2020fxf}. Nonetheless, the obtained bounds remain broadly consistent with those reported in \cite{Datta:2015aka, Datta:2016flx, Shaw:2019fin, Shaw:2020fxf}.

Additionally, for the sake of completeness, the six panels in Fig.\;\ref{blkt_bran} depict the variation of ${\text {Br}}(B^+\to K^+\nu\bar{\nu})$ for both zero and negative values of the BLT parameters. However, since the branching ratio curves corresponding to these combinations of BLT parameters do not intersect the upper limits (for both 95\% CL and 99\% CL) of the allowed experimental results, we are unable to establish lower limits for \(R^{-1}\) in these cases. Therefore, Table \ref{lower_R} provides the lower limit values of \(R^{-1}\) (for both 95\% CL and 99\% CL) corresponding to the combinations of BLT parameters where ${\text {Br}}(B^+\to K^+\nu\bar{\nu})$ intersects the 95\% and 99\% CL of the permissible range of the experimental value.

In this context, we wish to address the lower bounds on \( R^{-1} \) within the framework of UED, particularly in light of the current analysis concerning the branching ratio of the decay process \( B^+ \to K^+ \nu \bar{\nu} \). Our analysis yields UED results when the BLT parameters are set to zero, specifically for \( R_V = R_f = 0 \). Under these conditions, the KK-mass for the \( n^{th} \) KK-level is expressed as \( nR^{-1} \), while the overlap integrals \( I^n_1 \) and \( I^n_2 \) equal one. Consequently, the function \( X_n \) simplifies to its UED form. We have verified that when the BLT parameters are nullified, the expression for the function \( X_n \) aligns precisely with that presented in reference \cite{Buras:2002ej}\footnote{In the work by Buras et al. \cite{Buras:2002ej}, the authors did not account for any radiative corrections to the KK-masses in their analysis, resulting in the KK-mass for the \( n^{th} \) KK-mode being \( nR^{-1} \).}. However, our current analysis reveals a significant contrast from our previous findings \cite{Datta:2015aka, Datta:2016flx, Shaw:2019fin, Biswas:2017vhc, Shaw:2020fxf}. Specifically, for the case where \( R^{-1} \) is evaluated with \( R_V = R_f = 0 \), the branching ratio curve fails to intersect the upper limits (for both 95\% and 99\% CL) of the permissible experimental values. Consequently, we are unable to establish a lower limit for \( R^{-1} \) in the context of the branching ratio results for the decay process \( B^+ \to K^+ \nu \bar{\nu} \).

\begin{figure}[ht!]
\begin{center}
\includegraphics[scale=1.3,angle=0]{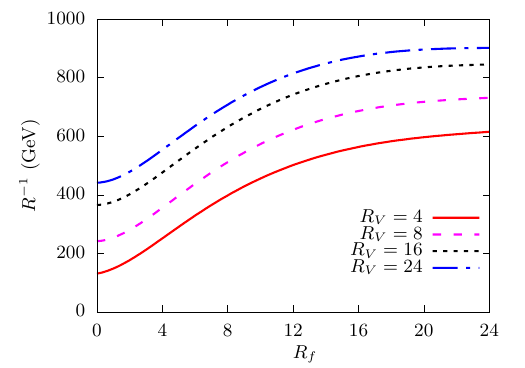}
\caption{This figure illustrates the exclusion contours derived from the branching ratio of the decay process \( B^+ \to K^+ \nu \bar{\nu} \) within the \( R_f - R^{-1} \) plane for four distinct values of \( R_V \). The exclusion curves are constructed based on the lower limit values of \( R^{-1} \), while incorporating contributions from up to 10 KK-levels in various loop functions necessary for the computation of WCs. The regions below each specific curve (corresponding to a fixed \( R_V \)) have been excluded based on the experimental measurements of the branching ratio with 99\% CL.}
\label{lowerR}
\end{center}
\end{figure}

Focusing on a specific panel that represents a positive value of \( R_V \), it is evident that both vanishing and negative BLT parameters do not impose any lower limit on \( R^{-1} \); thus, our analysis is confined to relevant positive BLT parameters. If we consider a vertical line corresponding to a fixed value of \( R^{-1} \), it becomes apparent that the branching ratio decreases as \( R_f \) increases. This trend is expected, as the masses of the KK-states diminish with increasing BLT parameters. Consequently, a lower KK-mass results in a enhanced branching ratio. Therefore, to achieve a constant branching ratio at elevated values of \( R^{-1} \), it is necessary to utilise higher values of \( R_f \). This effect is further amplified by increasing values of \( R_V \).

In Fig.\;\ref{lowerR}, we present the parameter space that has been excluded based on the experimentally determined branching ratio for the decay process \( B^+ \rightarrow K^+ \nu \bar{\nu} \). The figure illustrates contours corresponding to four distinct values of \( R_V \) within the \( R_f - R^{-1} \) plane. The region below every contour line is ruled out with 99\% CL. The characteristics of these contours can be elucidated by referring to Fig.\;\ref{blkt_bran}. 

\section{Summary}\label{concl}
In this study, we analyse the Kaluza-Klein contributions to the transitions \( B \to K^{(*)} \nu \bar{\nu} \) within a specific framework of (4+1)-dimensional Universal Extra Dimensions, wherein all Standard Model particles are permitted to propagate along a additional spatial dimension. This analysis is conducted in the context of boundary localised terms (BLTs), whose coefficients are parameterised to account for the unknown radiative corrections affecting the masses and couplings of the Kaluza-Klein modes. The inclusion of these boundary terms leads to nontrivial modifications of the masses and coupling strengths in the four-dimensional effective theory, deviating from the minimal formulation of the Universal Extra Dimensional model. By employing two distinct types of BLT parameters—namely, \( r_V \), which denotes the coefficients of boundary terms associated with the gauge and Higgs sectors, and \( r_f \), which characterises the coefficients of boundary terms related to fermions and Yukawa interactions—we investigate the \( B \to K^{(*)} \nu \bar{\nu} \) transitions within the Non-Minimal Universal Extra Dimensional framework.

The effective Hamiltonian governing the \( B \to K^{(*)} \nu \bar{\nu} \) transitions can be formulated in terms of four-fermion interactions, with the coefficients of these interactions being characterised by appropriate Wilson coefficients. Through the analysis of one-loop Feynman diagrams, as illustrated in Figs. \ref{pen}, \ref{self}, and \ref{box}, we have computed the coefficient associated with the operator responsible for the \( B \to K^{(*)} \nu \bar{\nu} \) transitions. Furthermore, by employing the Glashow-Iliopoulos-Maiani mechanism, we have incorporated contributions from all three generations of quarks into our evaluation. Additionally, in light of a recent study that connects the Higgs boson mass with the cut-off in a Universal Extra Dimensional framework \cite{Datta:2012db}, we have included contributions from up to 10 Kaluza-Klein modes in our calculations. The Kaluza-Klein contributions derived from the relevant one-loop Feynman diagrams have been combined with the corresponding contributions from the Standard Model (zero-mode) to provide a comprehensive analysis.

Following an assessment of the function \(X\), derived from one-loop Feynman diagrams, we have calculated the branching fraction for the decay \(B^+ \to K^+ \nu \bar{\nu}\) within the framework of the Non-Minimal Universal Extra Dimensional scenario. By comparing our theoretical predictions with the 99\% confidence level intervals of the experimentally allowed ranges \cite{Belle-II:2023esi}, we have successfully imposed constraints on the parameter space of the present version of the Non-Minimal Universal Extra Dimensional model. Additionally, we take into account the upper limit on the branching fraction for the decay \(B \to K^* \nu \bar{\nu}\) as reported by the Belle collaboration \cite{Belle:2017oht}, which serves as a significant constraint. This decay, occurring at the quark level, is governed by the same transition and characterised by the same Wilson coefficient as that involved in the decay \(B^+ \to K^+ \nu \bar{\nu}\).

It has been previously indicated that in the case of vanishing BLT parameters (i.e., \(R_f = R_V = 0\)), the outcomes align with those of the minimal version of the Universal Extra Dimensional scenario. Consequently, we have endeavoured to establish a lower bound on \(R^{-1}\) within the context of the Minimal Universal Extra Dimensional scenario under the condition of vanishing BLT parameters. However, our numerical analysis clearly demonstrates that the curve representing the branching ratio for the decay \(B^+ \to K^+ \nu \bar{\nu}\) (when \(R_f = R_V = 0\)) does not intersect with the upper limit of the experimental data. As a result, we are unable to ascertain a lower limit for \(R^{-1}\) in the  Universal Extra Dimensional scenario. This finding stands in stark contrast to our previous investigations in the realm of \(B\)-physics.

In the context of the Non-Minimal Universal Extra Dimensional scenario, the incorporation of various nonvanishing BLT parameters allows for an enhancement of the lower limit on \(R^{-1}\). For instance, by selecting a specific set of BLT parameters, namely \(R_V=8\) and \(R_f=16\), our analysis yields a lower limit of \(R^{-1} = 710.98\) GeV. Conversely, utilising another set of BLT parameters, \(R_V=24\) and \(R_f=24\), results in a lower limit of \(R^{-1} = 901.78\) GeV. These lower limit values are consistent with those derived from our previous investigations into rare decays of \(B\)-mesons \cite{Datta:2015aka, Datta:2016flx, Shaw:2019fin, Shaw:2020fxf}. Furthermore, it is noteworthy that the lower limits of \(R^{-1}\) for the same BLT parameter sets may experience a slight increase if the upper limit on the branching fraction for the decay \(B \to K^* \nu \bar{\nu}\), as reported by the Belle collaboration, is removed.
}

{\bf Acknowledgments} The author thanks Anindya Datta for insightful suggestions.
\begin{appendices}
\renewcommand{\thesection}{\Alph{section}}
\renewcommand{\theequation}{\thesection-\arabic{equation}} 

\setcounter{equation}{0}  

\section{The various Feynman diagrams which are contributing to the $B\to K^{(\ast)}\nu\bar{\nu}$}\label{appA}
\subsection{The \( Z \)-penguin and self-energy contributions to the functions \( C_n \), shown in Figs.~\ref{pen} and \ref{self}, are given below:}
The subsequent expressions, obtained from Fig.~\ref{pen}, correspond to the contributions of the \( Z \)-penguin diagrams to the functions \( C_n \):
\begin{eqnarray}
F_{1}(x_{f(n)}) \!\!\!\!&=&\!\!\!\!\frac{1}{8}\left(c^4_{fn}+s^4_{fn}-\frac{4}{3} s^2_W\right)
\left[ \Delta +\ln\frac{\mu^2}{M^2_{f^{(n)}}}-\frac{3}{2}+h_q\left(x_{f(n)}\right)-2 x_{f(n)} h_q\left(x_{f(n)}\right)\right](I^n_1)^2,\\
F_{2}(x_{f(n)}) \!\!\!\!&=& \!\!\!\!\frac{1}{4}c^2_{fn}s^2_{fn}\left[ \Delta +\ln\frac{\mu^2}{M^2_{f^{(n)}}}-\frac{3}{2}+h_q\left(x_{f(n)}\right)+2 x_{f(n)} h_q\left(x_{f(n)}\right)\right](I^n_1)^2,\\
F_{3}(x_{f(n)}) \!\!\!\!&=&\!\!\!\!\frac{1}{16 M^2_{W^{(n)}}}
  \Bigg[ \left( (m^{{(f)}}_{1})^2+ (m^{{(f)}}_{3})^2\right)\left(c_{f{n}}^{2} -\frac{4}{3} s^2_W \right)
    +\left((m^{{(f)}}_{2})^2+ (m^{{(f)}}_{4})^2\right)  \left(s_{fn}^2-\frac{4}{3} s^2_W \right)\Bigg]
\nonumber \\ &&
\times\left[ \Delta + 
      \ln \frac{\mu^2}{M^2_{f^{(n)}}}-\frac{1}{2}+h_q\left(x_{f(n)}\right) 
      -2 x_{f(n)} h_q\left(x_{f(n)}\right)\right],
\end{eqnarray}
\begin{eqnarray}      
F_{4}(x_{f(n)})\!\!\!\!&=& \!\!\!\!-\frac{1}{8 M^2_{W^{(n)}}} 
      \left( m^{{(f)}}_{1}\, m^{{(f)}}_{2}
    + m^{{(f)}}_{3}\, m^{{(f)}}_{4}\right)~c_{fn}~s_{fn} 
   \nonumber \\ &&
\times\left[ \Delta + 
  \ln\frac{\mu^2}{M^2_{f^{(n)}}}-\frac{1}{2}  +  h_q\left(x_{f(n)}\right) 
    + 2 x_{f(n)} h_q\left(x_{f(n)}\right)\right],\\
F_{5}(x_{f(n)})\!\!\!\!&=&\!\!\!\! -\frac{3}{4}c^2_W\left[\Delta + \ln \frac{\mu^2}{M^2_{W^{(n)}}}-\frac{1}{6}-x_{f(n)}h_w\left(x_{f(n)}\right)\right](I^n_1)^2,\\
F_{6}(x_{f(n)})\!\!\!\!&= &\!\!\!\!-\frac{1}{16\, M^{4}_{W^{(n)}}} 
  \Bigg[ \bigg( \left( 1-2\,s^2_W \right) M_W^2 + 2 c^2_W m^2_{V^{(n)}} \bigg) 
     \left((m^{{(f)}}_{1})^2+ (m^{{(f)}}_{2})^2\right)  
 \nonumber \\ && + \bigg(\left(1-2\,s^2_W \right) m^2_{V^{(n)}}   + 2 c^2_W M_W^2 \bigg)  \left((m^{{(f)}}_{3})^2+ (m^{{(f)}}_{4})^2\right)
        \Bigg]
 \nonumber \\ &&\times\left[ \Delta + 
  \ln \frac{\mu^2}{M_{W(n)}^2}+\frac{1}{2}  - x_{f(n)} h_w\left(x_{f(n)}\right) \right],
\\
F_{7}(x_{f(n)}) \!\!\!\!&=&\!\!\!\!\frac{M_W  m_{V^{(n)}}}{8 M^{4}_{W^{(n)}}}
 \left( m^{{(f)}}_{1} m^{{(f)}}_{3} +  m^{{(f)}}_{2}
  m^{{(f)}}_{4}\right) 
  \left[ \Delta + 
    \ln \frac{\mu^2}{M_{W(n)}^2}+\frac{1}{2}  - x_{f(n)}\, h_w\left(x_{f(n)}\right)\right],\\
F_{8}(x_{f(n)})\!\!\!&=&\!\!\!\frac{M_{f^{(n)}} }{2 M^{4}_{W^{(n)}}}
  \Bigg[ \bigg(s^2_W M_W^2 - c_W^2 m^2_{V^{(n)}}\bigg)\left(m^{{(f)}}_{1} ~c_{fn} +m^{{(f)}}_{2} ~s_{fn} \right)
\nonumber \\&& 
+ M_W
      m_{V^{(n)}}\left( m^{{(f)}}_{3} ~c_{fn} 
       + m^{{(f)}}_{4} ~s_{fn} \right)\Bigg]\,
       h_w\left(x_{f(n)}\right)I^n_1,
\end{eqnarray}
with the functions \( h_q \) and \( h_w \) defined as:
\begin{eqnarray}
    h_q(x) &=& \frac{1}{1-x} + \frac{\ln x}{(1-x)^2}\label{hq}, \\
    h_w(x) &=& \frac{1}{1-x} + \frac{x \ln x}{(1-x)^2}\label{hw} ~.
  \end{eqnarray}
The following expressions, derived from Fig.~\ref{self}, represent the self-energy diagram contributions to the functions \( C_n \):
{\small
 \begin{eqnarray}
     \Delta S_1\left(x_{f(n)} \right) \!\!\! &=&  \!\!\!\frac14 \left[\Delta
    -\frac{1}{2} \left\{\frac{1+x_{f(n)}}{1-x_{f(n)}} +  \frac{2
    x_{f(n)}^2 \ln x_{f(n)} }{(1-x_{f(n)})^2} \right\} - 
    \ln { \frac{M^2_{W^{(n)}}}{\mu^2}} \right](I^n_1)^2,\\
     \Delta S_2\left(x_{f(n)} \right)\! \!\! &=& \! \!\!\frac18 \left[
    (I^n_2)^2+ (I^n_1)^2x_f \right]  \left[\Delta +\frac{1}{2} 
\left\{ \frac{1-3x_{f(n)}}{1-x_{f(n)}} -
          \frac{2 x_{f(n)}^2\ln x_{f(n)} }{(1-x_{f(n)})^2} \right\}   - 
    \ln { \frac{M^2_{W^{(n)}}}{\mu^2}} \right] \!.
  \end{eqnarray}
}
 
Here, $\Delta=\frac{2}{\epsilon}+\ln 4\pi-\gamma_{E}$, \;\; $D=4-\epsilon$.

Note that the quantities \( I^n_1 \) and \( I^n_2 \), which appear in all the Feynman vertices, are the overlap integrals given in Eqs.~\ref{i1} and \ref{i2}, respectively~\cite{Datta:2015aka}:
\begin{equation}
I^n_1 = 2\sqrt{\frac{1+\frac{r_V}{\pi R}}{1+\frac{r_f}{\pi R}}}\left[ \frac{1}{\sqrt{1 + \frac{r^2_f m^2_{f^{(n)}}}{4} + \frac{r_f}{\pi R}}}\right]\left[ \frac{1}{\sqrt{1 + \frac{r^2_V m^2_{V^{(n)}}}{4} + \frac{r_V}{\pi R}}}\right]\frac{m^2_{V^{(n)}}}{\left(m^2_{V^{(n)}} - m^2_{f^{(n)}}\right)}\frac{\left(r_{f} - r_{V}\right)}{\pi R},
\label{i1}
\end{equation}

\begin{equation}
I^n_2 = 2\sqrt{\frac{1+\frac{r_V}{\pi R}}{1+\frac{r_f}{\pi R}}}\left[ \frac{1}{\sqrt{1 + \frac{r^2_f m^2_{f^{(n)}}}{4} + \frac{r_f}{\pi R}}}\right]\left[ \frac{1}{\sqrt{1 + \frac{r^2_V m^2_{V^{(n)}}}{4} + \frac{r_V}{\pi R}}}\right]\frac{m_{V^{(n)}}m_{f^{(n)}}}{\left(m^2_{V^{(n)}} - m^2_{f^{(n)}}\right)}\frac{\left(r_{f} - r_{V}\right)}{\pi R}.
\label{i2}
\end{equation}

\subsection{The box diagram contributions to the functions \( B^{\nu\bar{\nu}}_n \), as depicted in Fig.~\ref{box}, are given by:}
\begin{eqnarray}
    G_{WW(n)} &=&-\frac{M_W^2}{M^2_{W^{(n)}}}~ U(x_{f(n)}, x_{e(n)})(I^n_1)^4,\\
    G_{WG(n)} &=& \frac12\,\frac{M_W^2 M_{f^{(n)}}m_{e^{(n)}}}{{M^6_{W^{(n)}}}} \left[
      m_1^{(f)} c_{f(n)} + m_2^{(f)} s_{f(n)} \right] ~m_1^{(e)}~
      \widetilde U(x_{f(n)}, x_{e(n)})(I^n_1)^2~, \\
       G_{WH(n)} &=& \frac12\,\frac{M_W^2 M_{f^{(n)}}m_{e^{(n)}}}{{M^6_{W^{(n)}}}} \left[
      m_3^{(f)} c_{f(n)} + m_4^{(f)} s_{f(n)} \right]~ m_3^{(e)} ~
      \widetilde U(x_{f(n)}, x_{e(n)})(I^n_1)^2~, \\
      G_{GH(n)} &=& -\frac{1}{8} \frac{M_W^2}{{M^6_{W^{(n)}}}}
    \left[m_1^{(f)} m_3^{(f)} + m_2^{(f)}
      m_4^{(f)}\right]~m_1^{(e)}\, m_3^{(e)}~ U(x_{f(n)}, x_{e(n)})~,\\
     G_{GG(n)} &=&-\frac{1}{16} \frac{M_W^2}{{M^6_{W^{(n)}}}}
   \left[(m_1^{(f)})^2 + (m_2^{(f)})^2 \right] (m_1^{(e)})^2~  U(x_{f(n)}, x_{e(n)})~,\\
    G_{HH(n)} &=& -\frac{1}{16}\frac{M_W^2}{{M^6_{W^{(n)}}}}
   \left[(m_3^{(f)})^2 + (m_4^{(f)})^2\right] (m_3^{(e)})^2 ~  
U(x_{f(n)}, x_{e(n)})~.
 \end{eqnarray}
The definitions of the functions \( U \) and \( \widetilde{U} \) are given by:
\begin{eqnarray}\label{U1}
    U (x_t, x_u) &=&\frac{x_t^2\log{x_t}}{(x_t-x_u)(1-x_t)^2} +
    \frac{x_u^2\log{x_u}}{(x_u-x_t)(1-x_u)^2} + \frac{1}{(1-x_u)(1-x_t)}~,\\ 
  \label{U3}
    \widetilde U (x_t, x_u) &=& \frac{x_t\log{x_t}}{(x_t-x_u)(1-x_t)^2} +
    \frac{x_u\log{x_u}}{(x_u-x_t)(1-x_u)^2} + \frac{1}{(1-x_u)(1-x_t)}~.
\end{eqnarray}
\setcounter{equation}{0} 
\section{Feynman Rules for \boldmath{$B\to K^{(\ast)}\nu\bar{\nu}$} in the NMUED Framework
}\label{fyerul}
\noindent The relevant Feynman rules for various interaction vertices are listed below, assuming all momenta and fields to be incoming. While similar Feynman rules are discussed in~\cite{Buras:2002ej}, they have been appropriately modified in the context of the NMUED model, as presented in~\cite{Datta:2015aka}.
\begin{enumerate}
\item $Z^{\mu}W^{\nu\pm}S^{\mp}$
$\displaystyle : \frac{g_2}{c_W M_{W^{(n)}}} g_{\mu\nu} C$, where $C$ is expressed as follows:

\begin{equation}
 \begin{aligned}
  Z^{\mu} W^{(n)+} G^{(n)-}: C   &= - s^2_W M_W^2 + c^2_W m^2_{V^{(n)}},\\ 
  Z^{\mu} W^{(n)-} G^{(n)+}: C   &= s^2_W M_W^2 - c^2_W m^2_{V^{(n)}},\\
  Z^{\mu} W^{(n)+} H^{(n)-}: C   &= -iM_W m_{V^{(n)}},\\
  Z^{\mu} W^{(n)-} H^{(n)+}: C   &= iM_W m_{V^{(n)}}.  
 \end{aligned}
\end{equation}

\item $Z^{\mu}S^{\pm}_1S^{\mp}_2$
$\displaystyle : \frac{ig_2}{2c_W M^2_{W^{(n)}}} (k_2-k_1)_{\mu} C$, where $C$ is expressed as follows:

\begin{equation}
 \begin{aligned}
  Z^{\mu} G^{(n)+} G^{(n)-}: C &= -\left(c^2_W-s^2_W\right) M_W^2 - 2c^2_W m^2_{V^{(n)}},\\   
  Z^{\mu} H^{(n)+} H^{(n)-}: C &=  - 2c^2_W M_W^2 -  \left(c^2_W-s^2_W\right)m^2_{V^{(n)}},\\
  Z^{\mu} G^{(n)+} H^{(n)-}: C &= iM_W m_{V^{(n)}},\\ 
  Z^{\mu} G^{(n)-} H^{(n)+}: C &= -iM_W m_{V^{(n)}}.
\end{aligned}
\end{equation}

In this context, the scalar fields are presented as $S\equiv H,G.$

\item $Z^{\mu}(k_1)W^{\nu+}(k_2)W^{\lambda-}(k_3)$
$\displaystyle :$
\begin{equation}
 ig_2c_W \left[ g_{\mu\nu} (k_2
      -k_1)_\lambda + g_{\mu\lambda} (k_1 -k_3)_\nu +
    g_{\lambda\nu} (k_3 -k_2)_\mu \right].
\end{equation}

\item $Z^{\mu}{\overline{f}_1} f_2$
  $\displaystyle  : \frac{i g_2}{6 c_W} \gamma_\mu \left( P_L C_L +
    P_R C_R \right)$, where $C_L$ and $C_R$ are expressed as follows:

\begin{equation}
\begin{aligned}
  & Z^{\mu} \bar{u_i} u_i:  &
  &\left\{\begin{array}{l}C_L = 3-4s^2_W,\\
      C_R = -4s^2_W,\end{array}\right.
  && Z^{\mu} \bar{d_j} d_j:    & 
  &\left\{\begin{array}{l}C_L = -3+2s^2_W,\\
      C_R = 2s^2_W,\end{array}\right.\\
  &Z^{\mu} \bar {\nu_i} \nu_i:  & 
  &\left\{\begin{array}{l}C_L = 3,\\
      C_R = 0,\end{array}\right.
  && Z^{\mu} \bar {e_j} e_j:     &
  &\left\{\begin{array}{l}C_L = -3+6s^2_W,\\
      C_R = 6s^2_W ,\end{array}\right.\\
  & Z^{\mu} {\overline{T}^{1(n)}_i} T^{1(n)}_i: &   
  &\left\{\begin{array}{l}C_L = -4s^2_W+3c^2_{in},\\
      C_R = -4s^2_W+3c^2_{in},\end{array}\right.
  && Z^{\mu} {\overline{T}^{2(n)}_i} T^{2(n)}_i:     &
  &\left\{\begin{array}{l}C_L = -4s^2_W+3s^2_{in},\\
      C_R = -4s^2_W+3s^2_{in},\end{array}\right.\\
  & Z^{\mu} {\overline{T}^{1(n)}_i} T^{2(n)}_i:  &    
  &\left\{\begin{array}{l}C_L = -3s_{in} c_{in},\\
      C_R = 3s_{in} c_{in},\end{array}\right.
  &&Z^{\mu} {\overline{T}^{2(n)}_i} T^{1(n)}_i:     & 
  &\left\{\begin{array}{l}C_L = -3s_{in} c_{in},\\
      C_R = 3s_{in} c_{in}.\end{array}\right.
\end{aligned}
\end{equation}

\item $S^{\pm}{\overline{f}_1} f_2$
  $\displaystyle  = \frac{g_2}{\sqrt{2} M_{W^{(n)}}} (P_L C_L + P_R C_R)$, where $C_L$ and $C_R$ are expressed as follows:

\begin{equation}
\begin{aligned}
  & G^+ \bar{u_i} d_j :  &
  &\left\{\begin{array}{l}C_L = -m_i V_{ij},\\
      C_R = m_j V_{ij},\end{array}\right.
  &&G^- \bar{d_j} u_i :     &
  &\left\{\begin{array}{l}C_L = -m_j V_{ij}^*,\\
      C_R = m_i V_{ij}^*,\end{array}\right.\\
  & G^{(n)+}{\overline{T}^{1(n)}_i} d_j :  &
  &\left\{\begin{array}{l}C_L = -m_1^{(i)} V_{ij},\\
      C_R = M_1^{(i,j)} V_{ij},\end{array}\right.
  &&G^{(n)-}\bar{d_j}T^{1(n)}_i :   &
  &\left\{\begin{array}{l}C_L = -M_1^{(i,j)} V_{ij}^*,\\
     C_R = m_1^{(i)} V_{ij}^*,\end{array}\right.\\
  & G^{(n)+}{\overline{T}^{2(n)}_i} d_j :  &
  &\left\{\begin{array}{l}C_L = m_2^{(i)} V_{ij},\\
      C_R =-M_2^{(i,j)} V_{ij},\end{array}\right.
  &&G^{(n)-}\bar{d_j}T^{2(n)}_i :   &
  &\left\{\begin{array}{l}C_L = M_2^{(i,j)} V_{ij}^*,\\
     C_R =-m_2^{(i)} V_{ij}^*,\end{array}\right.\\
  & G^+ \bar{\nu_i} e_j :  &
  &\left\{\begin{array}{l}C_L = 0,\\
      C_R = m_j \delta_{ij},\end{array}\right.
  &&G^- \bar{e_j} \nu_i :     &
  &\left\{\begin{array}{l}C_L = -m_j \delta_{ij},\\
      C_R = 0,\end{array}\right.\\
  & G^{(n)+} \bar{\nu_i} {\mathcal L}^{(n)}_j :  &
  &\left\{\begin{array}{l}C_L = 0,\\
      C_R = m^{(j)}_1 \delta_{ij},\end{array}\right.
  &&G^{(n)-}{\overline{\mathcal L}^{(n)}_j} \nu_i :     &
  &\left\{\begin{array}{l}C_L = -m^{(j)}_1 \delta_{ij},\\
      C_R = 0,\end{array}\right.\\
  & G^{(n)+} \bar{\nu_i} {\mathcal E}^{(n)}_j :  &
  &\left\{\begin{array}{l}C_L = 0,\\
      C_R =-m^{(j)}_2 \delta_{ij},\end{array}\right.
  &&G^{(n)-}{\overline{\mathcal E}^{(n)}_j} \nu_i :     &
  &\left\{\begin{array}{l}C_L = m^{(j)}_2 \delta_{ij},\\
      C_R = 0,\end{array}\right.\\
  & H^{(n)+}{\overline{T}^{1(n)}_i} d_j :  &
  &\left\{\begin{array}{l}C_L = -m_3^{(i)} V_{ij},\\
      C_R = M_3^{(i,j)} V_{ij},\end{array}\right.
  &&H^{(n)-}\bar{d_j}T^{1(n)}_i :   &
  &\left\{\begin{array}{l}C_L = -M_3^{(i,j)} V_{ij}^*,\\
     C_R = m_3^{(i)} V_{ij}^*,\end{array}\right.\\
  & H^{(n)+}{\overline{T}^{2(n)}_i} d_j :  &
  &\left\{\begin{array}{l}C_L = m_4^{(i)} V_{ij},\\
      C_R =-M_4^{(i,j)} V_{ij},\end{array}\right.
  &&H^{(n)-}\bar{d_j}T^{2(n)}_i :   &
  &\left\{\begin{array}{l}C_L = M_4^{(i,j)} V_{ij}^*,\\
     C_R =-m_4^{(i)} V_{ij}^*,\end{array}\right.\\
  & H^{(n)+} \bar{\nu_i} {\mathcal L}^{(n)}_j :  &
  &\left\{\begin{array}{l}C_L = 0,\\
      C_R = m^{(j)}_3 \delta_{ij},\end{array}\right.
  &&H^{(n)-}{\overline{\mathcal L}^{(n)}_j} \nu_i :     &
  &\left\{\begin{array}{l}C_L = -m^{(j)}_3 \delta_{ij},\\
      C_R = 0,\end{array}\right.\\
  & H^{(n)+} \bar{\nu_i} {\mathcal E}^{(n)}_j :  &
  &\left\{\begin{array}{l}C_L = 0,\\
      C_R =-m^{(j)}_4 \delta_{ij},\end{array}\right.
  &&H^{(n)-}{\overline{\mathcal E}^{(n)}_j} \nu_i :     &
  &\left\{\begin{array}{l}C_L = m^{(j)}_4 \delta_{ij},\\
      C_R = 0,\end{array}\right.\\
\end{aligned}
\end{equation}

In this context, the fermion fields are presented as $f\equiv u, d, T^1_t, T^2_t, \nu, e, {\mathcal L}_{\nu}, {\mathcal L}_e, {\mathcal E}.$

\item $W^{\mu\pm}{\overline{f}_1}f_2$
  $\displaystyle  :  \frac{i g_2}{\sqrt{2}} \gamma_\mu P_L C_L$, where $C_L$ is expressed as follows:

\begin{equation}
\begin{aligned}
  & W^{\mu+}\bar{u_i} d_j : &&     C_L = V_{ij},
  && W^{\mu-}\bar{d_j} u_i : &&    C_L = V^*_{ij},\\
  & W^{\mu(n)+}{\overline{T}^{1(n)}_i}d_j : &&   C_L = I^n_1\;c_{in} V_{ij},
  &&W^{\mu(n)-}\bar{d_j}{{T}^{1(n)}_i} : && C_L = I^n_1\;c_{in} V^*_{ij},\\
  & W^{\mu(n)+}{\overline{T}^{2(n)}_i}d_j : &&   C_L = -I^n_1\;s_{in} V_{ij},
  &&W^{\mu(n)-}\bar{d_j}{{T}^{2(n)}_i} : && C_L = -I^n_1\;s_{in}V^*_{ij},\\
  & W^{\mu+}\bar {\nu_i} e_j : &&     C_L = \delta_{ij},
  && W^{\mu-}\bar{e_j} \nu_i : &&    C_L = \delta_{ij},\\
  & W^{\mu(n)+}\bar {\nu_i} {\mathcal L}^{(n)}_j : &&     C_L = I^n_1\;\delta_{ij},
  &\hspace{12ex} & W^{\mu(n)-}{\overline{\mathcal L}^{(n)}_j} \nu_i : &&    C_L =  I^n_1\;\delta_{ij},\\
  & W^{\mu(n)+}\bar {\nu_i} {\mathcal E}^{(n)}_j : &&     C_L = 0,
  && W^{\mu(n)-}{\overline{\mathcal E}^{(n)}_j} \nu_i : &&    C_L = 0.
\end{aligned}
\end{equation}
\end{enumerate}
The mass parameters $m_x^{(i)}$ are defined by the following expressions:
\begin{equation}
\label{mparameters}
  \begin{aligned}
    m_1^{(i)} &= I^n_2\;m_{V^{(n)}}c_{in} +I^n_1\;m_i s_{in},\\
    m_2^{(i)} &= -I^n_2\;m_{V^{(n)}}s_{in}+I^n_1\;m_i c_{in},\\
    m_3^{(i)} &= -I^n_2\;iM_W c_{in} +I^n_1\;i\frac{m_{V^{(n)}}m_i}{M_W}s_{in},\\
    m_4^{(i)} &= I^n_2\;iM_W s_{in}+I^n_1\;i\frac{m_{V^{(n)}}m_i}{M_W}c_{in}.
  \end{aligned}
\end{equation}
\noindent Here, \( m_i \) denotes the mass of the zero-mode \textit{up-type} fermion, while \( c_{in} = \cos(\alpha_{in}) \) and \( s_{in} = \sin(\alpha_{in}) \), with \( \alpha_{in} \) defined earlier.

The mass parameters $M_x^{(i,j)}$ are given by the following expressions:
\begin{equation}\label{Mparameters}
  \begin{aligned}
    M_1^{(i,j)}  &= I^n_1\;m_j c_{in},\\
    M_2^{(i,j)}  &= I^n_1\;m_j s_{in},\\
    M_3^{(i,j)}  &= I^n_1\;i\frac{m_{V^{(n)}}m_j}{M_W}c_{in},\\
    M_4^{(i,j)}  &= I^n_1\;i\frac{m_{V^{(n)}}m_j}{M_W}s_{in}.
  \end{aligned}
\end{equation}
\noindent where \( m_j \) denotes the mass of the zero-mode \textit{down-type} fermion.

\setcounter{equation}{0} 
\section{Form Factors relevant to the \boldmath{$B \to K^{(*)} \nu \bar{\nu}$} transitions}\label{BFF}
This appendix provides a brief overview of the \( B \to K^{(*)} \) form factors relevant to the rare \( B \)-meson decays studied herein.
\subsection{\boldmath Details of the Form Factors for \boldmath{$B^+ \to K^+ \nu \bar{\nu}$} transition}\label{BtoKst:ff}
The long-range effects associated with the hadronic dynamics of the decay process \( B^+ \to K^+ \nu \bar{\nu} \) are characterised by the subsequent matrix elements \cite{Ball:2004ye, Bartsch:2009qp}:
\begin{eqnarray}
\langle K^+(p')|\bar s\gamma^\mu b|B^+(p)\rangle
&=& f_+(s)\, (p+p')^\mu +[f_0(s)-f_+(s)]\,\frac{m^2_B-m^2_K}{q^2}q^\mu\;.
\label{fpf0def}
\end{eqnarray}
In this context, the relevant form factors are denoted as \( f_+ \) and \( f_0 \). Additionally, we define \( q = p - p' \) and \( s = \frac{q^2}{m_B^2} \). It is noteworthy that the contributions from the \( f_0 \) term are negligible in the expression for the differential branching ratio (refer to Eq.~\ref{brBtoKnunubar}), primarily due to the extremely small masses of the neutrinos. The updated parameterisation of the form factors for the \( B^+ \to K^+ \) transition, as presented by the FLAG collaboration~\cite{FlavourLatticeAveragingGroupFLAG:2021npn}, is available for reference.

\subsection{Details of form factors for $B\to K^{*}\nu{\bar{\nu}}$ transitions}\label{BtoKst:ff}
The matrix elements for the relevant operators for $B(p)\to K^{*}(k)$ transitions interms of momentum transfer ($q=p-k$) dependent form factors can be written as \cite{Altmannshofer:2008dz, Bharucha:2015bzk}:
\begin{align}
\label{eq:BKst-ff}
\langle {K}^{\ast}(k)|\bar{s}\gamma_\mu(1&-\gamma_5) b|{B}(p) \rangle = \epsilon_{\mu\nu\rho\sigma} \epsilon^{\ast\nu}p^\rho k^\sigma \dfrac{2 V(q^2)}{m_B+m_{K^\ast}}
-i\epsilon_\mu^\ast(m_B+m_{K^\ast})A_1(q^2) \nonumber\\
&+ i(2p-q)_\mu (\epsilon^\ast\cdot q)\dfrac{A_2(q^2)}{m_B+m_{K^\ast}}
+i q_\mu (\epsilon^\ast\cdot q) \dfrac{2 m_{K^\ast}}{q^2} \left[A_3(q^2)-A_0(q^2)\right]\;. 
\end{align}
In this context, $\epsilon_\mu$ denotes the polarisation vector of the $K^\ast$ meson, while $V(q^2)$ and $A_{0,1,2,3}(q^2)$ represent the associated form factors. The definition provided utilises $A_3(q^2)$, whereas Eq.\;\ref{Ffunc} employs $A_{12}(q^2)$. Both of these form factors are interconnected with $A_1(q^2)$ and $A_2(q^2)$ through the following relationships:
\begin{align}
A_3(q^2) =& \dfrac{m_B+m_{K^\ast}}{2 m_{K^\ast}}A_1(q^2)
- \dfrac{m_B-m_{K^\ast}}{2 m_{K^\ast}}A_2(q^2)\;, \\
A_{12}(q^2) =& \dfrac{(m_B+m_{K^\ast}) (m_B^2 -m_{K^\ast}^2 -q^2)}{16 m_B m_{K^\ast}^2} A_1(q^2) 
- \dfrac{\lambda_{K^\ast}}{16 m_B m_{K^\ast}^2 (m_B+m_{K^\ast}) } A_2(q^2)\;. 
\end{align}
At the specific point of $q^2=0$, these form factors satisfy the relations: $8 m_B m_{K^\ast} A_{12}(0) = (m_B^2 - m_{K^\ast}^2) A_0(0)$ and $A_3(0)=A_0(0)$. It is noteworthy that the pseudoscalar form factor $A_0(q^2)$ does not influence the decay rate in the limit of massless neutrinos. Consequently, three independent form factors—namely $V$, $A_{1}$, and $A_{2}$—are essential for the computation of the decay process $B\to K^\ast\nu\bar{\nu}$. An updated parameterisation of the form factors pertinent to the \( B \to K^\ast \) transition can be found in the reference \cite{Bharucha:2015bzk}.
\end{appendices}

\providecommand{\href}[2]{#2}\begingroup\raggedright\endgroup


\begin{thebibliography}{100}

\bibitem{Aad:2012tfa}
{\scshape ATLAS} collaboration, G.~Aad et~al., \emph{{Observation of a new
  particle in the search for the Standard Model Higgs boson with the ATLAS
  detector at the LHC}},
  \href{https://doi.org/10.1016/j.physletb.2012.08.020}{\emph{Phys. Lett. B}
  {\bfseries 716} (2012) 1--29},
  [\href{https://arxiv.org/abs/1207.7214}{{\ttfamily 1207.7214}}].

\bibitem{Chatrchyan:2012xdj}
{\scshape CMS} collaboration, S.~Chatrchyan et~al., \emph{{Observation of a New
  Boson at a Mass of 125 GeV with the CMS Experiment at the LHC}},
  \href{https://doi.org/10.1016/j.physletb.2012.08.021}{\emph{Phys. Lett. B}
  {\bfseries 716} (2012) 30--61},
  [\href{https://arxiv.org/abs/1207.7235}{{\ttfamily 1207.7235}}].

\bibitem{Belle-II:2023esi}
{\scshape Belle-II} collaboration, I.~Adachi et~al., \emph{{Evidence for
  B+{\textrightarrow}K+{\ensuremath{\nu}}{\ensuremath{\nu}}{\textasciimacron}
  decays}}, \href{https://doi.org/10.1103/PhysRevD.109.112006}{\emph{Phys. Rev.
  D} {\bfseries 109} (2024) 112006},
  [\href{https://arxiv.org/abs/2311.14647}{{\ttfamily 2311.14647}}].

\bibitem{Allwicher:2023xba}
L.~Allwicher, D.~Becirevic, G.~Piazza, S.~Rosauro-Alcaraz and O.~Sumensari,
  \emph{{Understanding the first measurement of
  B(B{\textrightarrow}K{\ensuremath{\nu}}{\ensuremath{\nu}}{\textasciimacron})}},
  \href{https://doi.org/10.1016/j.physletb.2023.138411}{\emph{Phys. Lett. B}
  {\bfseries 848} (2024) 138411},
  [\href{https://arxiv.org/abs/2309.02246}{{\ttfamily 2309.02246}}].

\bibitem{Bause:2023mfe}
R.~Bause, H.~Gisbert and G.~Hiller, \emph{{Implications of an enhanced
  B{\textrightarrow}K{\ensuremath{\nu}}{\ensuremath{\nu}}{\textasciimacron}
  branching ratio}},
  \href{https://doi.org/10.1103/PhysRevD.109.015006}{\emph{Phys. Rev. D}
  {\bfseries 109} (2024) 015006},
  [\href{https://arxiv.org/abs/2309.00075}{{\ttfamily 2309.00075}}].

\bibitem{Altmannshofer:2023hkn}
W.~Altmannshofer, A.~Crivellin, H.~Haigh, G.~Inguglia and J.~Martin~Camalich,
  \emph{{Light new physics in
  B{\textrightarrow}K(*){\ensuremath{\nu}}{\ensuremath{\nu}}{\textasciimacron}?}},
  \href{https://doi.org/10.1103/PhysRevD.109.075008}{\emph{Phys. Rev. D}
  {\bfseries 109} (2024) 075008},
  [\href{https://arxiv.org/abs/2311.14629}{{\ttfamily 2311.14629}}].

\bibitem{McKeen:2023uzo}
D.~McKeen, J.~N. Ng and D.~Tuckler, \emph{{Higgs portal interpretation of the
  Belle II B+{\textrightarrow}K+{\ensuremath{\nu}}{\ensuremath{\nu}}
  measurement}}, \href{https://doi.org/10.1103/PhysRevD.109.075006}{\emph{Phys.
  Rev. D} {\bfseries 109} (2024) 075006},
  [\href{https://arxiv.org/abs/2312.00982}{{\ttfamily 2312.00982}}].

\bibitem{Fridell:2023ssf}
K.~Fridell, M.~Ghosh, T.~Okui and K.~Tobioka, \emph{{Decoding the
  B{\textrightarrow}K{\ensuremath{\nu}}{\ensuremath{\nu}} excess at Belle II:
  Kinematics, operators, and masses}},
  \href{https://doi.org/10.1103/PhysRevD.109.115006}{\emph{Phys. Rev. D}
  {\bfseries 109} (2024) 115006},
  [\href{https://arxiv.org/abs/2312.12507}{{\ttfamily 2312.12507}}].

\bibitem{Gabrielli:2024wys}
E.~Gabrielli, L.~Marzola, K.~M{\"u}{\"u}rsepp and M.~Raidal, \emph{{Explaining
  the $B^+\rightarrow K^+ \nu \bar{\nu }$ excess via a massless dark photon}},
  \href{https://doi.org/10.1140/epjc/s10052-024-12818-2}{\emph{Eur. Phys. J. C}
  {\bfseries 84} (2024) 460},
  [\href{https://arxiv.org/abs/2402.05901}{{\ttfamily 2402.05901}}].

\bibitem{Felkl:2023ayn}
T.~Felkl, A.~Giri, R.~Mohanta and M.~A. Schmidt, \emph{{When energy goes
  missing: new physics in $b\rightarrow s \nu \nu $ with sterile neutrinos}},
  \href{https://doi.org/10.1140/epjc/s10052-023-12326-9}{\emph{Eur. Phys. J. C}
  {\bfseries 83} (2023) 1135},
  [\href{https://arxiv.org/abs/2309.02940}{{\ttfamily 2309.02940}}].

\bibitem{Wang:2023trd}
Z.~S. Wang, H.~K. Dreiner and J.~Y. G{\"u}nther, \emph{{The decay $B\rightarrow
  K+\nu +\bar{\nu }$ at Belle II and a massless bino in R-parity-violating
  supersymmetry}},
  \href{https://doi.org/10.1140/epjc/s10052-025-13745-6}{\emph{Eur. Phys. J. C}
  {\bfseries 85} (2025) 66},
  [\href{https://arxiv.org/abs/2309.03727}{{\ttfamily 2309.03727}}].

\bibitem{He:2023bnk}
X.-G. He, X.-D. Ma and G.~Valencia, \emph{{Revisiting models that enhance
  B+{\textrightarrow}K+{\ensuremath{\nu}}{\ensuremath{\nu}}{\textasciimacron}
  in light of the new Belle II measurement}},
  \href{https://doi.org/10.1103/PhysRevD.109.075019}{\emph{Phys. Rev. D}
  {\bfseries 109} (2024) 075019},
  [\href{https://arxiv.org/abs/2309.12741}{{\ttfamily 2309.12741}}].

\bibitem{Datta:2023iln}
A.~Datta, D.~Marfatia and L.~Mukherjee,
  \emph{{B{\textrightarrow}K{\ensuremath{\nu}}{\ensuremath{\nu}}{\textasciimacron},
  MiniBooNE and muon g-2 anomalies from a dark sector}},
  \href{https://doi.org/10.1103/PhysRevD.109.L031701}{\emph{Phys. Rev. D}
  {\bfseries 109} (2024) L031701},
  [\href{https://arxiv.org/abs/2310.15136}{{\ttfamily 2310.15136}}].

\bibitem{Ho:2024cwk}
S.-Y. Ho, J.~Kim and P.~Ko, \emph{{Recent
  B+{\textrightarrow}K+{\ensuremath{\nu}}{\ensuremath{\nu}}{\textasciimacron}
  excess and muon g-2 illuminating light dark sector with Higgs portal}},
  \href{https://doi.org/10.1103/PhysRevD.111.055029}{\emph{Phys. Rev. D}
  {\bfseries 111} (2025) 055029},
  [\href{https://arxiv.org/abs/2401.10112}{{\ttfamily 2401.10112}}].

\bibitem{Chen:2024jlj}
F.-Z. Chen, Q.~Wen and F.~Xu, \emph{{Correlating $B\rightarrow K^{(*)} \nu
  \bar{\nu }$ and flavor anomalies in SMEFT}},
  \href{https://doi.org/10.1140/epjc/s10052-024-13425-x}{\emph{Eur. Phys. J. C}
  {\bfseries 84} (2024) 1012},
  [\href{https://arxiv.org/abs/2401.11552}{{\ttfamily 2401.11552}}].

\bibitem{Hou:2024vyw}
B.-F. Hou, X.-Q. Li, M.~Shen, Y.-D. Yang and X.-B. Yuan, \emph{{Deciphering the
  Belle II data on $ B\to K\nu \overline{\nu} $ decay in the (dark) SMEFT with
  minimal flavour violation}},
  \href{https://doi.org/10.1007/JHEP06(2024)172}{\emph{JHEP} {\bfseries 06}
  (2024) 172}, [\href{https://arxiv.org/abs/2402.19208}{{\ttfamily
  2402.19208}}].

\bibitem{He:2024iju}
X.-G. He, X.-D. Ma, M.~A. Schmidt, G.~Valencia and R.~R. Volkas, \emph{{Scalar
  dark matter explanation of the excess in the Belle II
  B$^{+}${\textrightarrow} K$^{+}$+ invisible measurement}},
  \href{https://doi.org/10.1007/JHEP07(2024)168}{\emph{JHEP} {\bfseries 07}
  (2024) 168}, [\href{https://arxiv.org/abs/2403.12485}{{\ttfamily
  2403.12485}}].

\bibitem{Bolton:2024egx}
P.~D. Bolton, S.~Fajfer, J.~F. Kamenik and M.~Novoa-Brunet, \emph{{Signatures
  of light new particles in B{\textrightarrow}K(*)Emiss}},
  \href{https://doi.org/10.1103/PhysRevD.110.055001}{\emph{Phys. Rev. D}
  {\bfseries 110} (2024) 055001},
  [\href{https://arxiv.org/abs/2403.13887}{{\ttfamily 2403.13887}}].

\bibitem{Buras:2024ewl}
A.~J. Buras, J.~Harz and M.~A. Mojahed, \emph{{Disentangling new physics in $
  K\to \pi \nu \overline{\nu} $ and $ B\to K\left({K}^{\ast}\right)\nu
  \overline{\nu} $ observables}},
  \href{https://doi.org/10.1007/JHEP10(2024)087}{\emph{JHEP} {\bfseries 10}
  (2024) 087}, [\href{https://arxiv.org/abs/2405.06742}{{\ttfamily
  2405.06742}}].

\bibitem{Altmannshofer:2024kxb}
W.~Altmannshofer and S.~Roy, \emph{{Joint explanation of the
  B{\textrightarrow}{\ensuremath{\pi}}K puzzle and the
  B{\textrightarrow}K{\ensuremath{\nu}}{\ensuremath{\nu}}{\textasciimacron}
  excess}}, \href{https://doi.org/10.1103/PhysRevD.111.075029}{\emph{Phys. Rev.
  D} {\bfseries 111} (2025) 075029},
  [\href{https://arxiv.org/abs/2411.06592}{{\ttfamily 2411.06592}}].

\bibitem{Marzocca:2024hua}
D.~Marzocca, M.~Nardecchia, A.~Stanzione and C.~Toni, \emph{{Implications of $B
  \rightarrow K \nu {\bar{\nu }}$ under rank-one flavor violation hypothesis}},
  \href{https://doi.org/10.1140/epjc/s10052-024-13534-7}{\emph{Eur. Phys. J. C}
  {\bfseries 84} (2024) 1217},
  [\href{https://arxiv.org/abs/2404.06533}{{\ttfamily 2404.06533}}].

\bibitem{Hu:2024mgf}
Q.-Y. Hu, \emph{{Are the new particles heavy or light in $b \rightarrow s
  E_{\textrm{miss}}$?}},
  \href{https://doi.org/10.1140/epjc/s10052-025-14290-y}{\emph{Eur. Phys. J. C}
  {\bfseries 85} (2025) 556},
  [\href{https://arxiv.org/abs/2412.19084}{{\ttfamily 2412.19084}}].

\bibitem{Allwicher:2024ncl}
L.~Allwicher, M.~Bordone, G.~Isidori, G.~Piazza and A.~Stanzione,
  \emph{{Probing third-generation New Physics with
  K{\textrightarrow}{\ensuremath{\pi}}{\ensuremath{\nu}}{\ensuremath{\nu}}{\textasciimacron}
  and
  B{\textrightarrow}K({\textasteriskcentered}){\ensuremath{\nu}}{\ensuremath{\nu}}{\textasciimacron}}},
  \href{https://doi.org/10.1016/j.physletb.2025.139295}{\emph{Phys. Lett. B}
  {\bfseries 861} (2025) 139295},
  [\href{https://arxiv.org/abs/2410.21444}{{\ttfamily 2410.21444}}].

\bibitem{Berezhnoy:2023rxx}
A.~Berezhnoy and D.~Melikhov, \emph{{$B\to K^* M_X$ vs $B\to K M_X$ as a probe
  of a scalar-mediator dark matter scenario}},
  \href{https://doi.org/10.1209/0295-5075/ad1d03}{\emph{EPL} {\bfseries 145}
  (2024) 14001}, [\href{https://arxiv.org/abs/2309.17191}{{\ttfamily
  2309.17191}}].

\bibitem{Calibbi:2025rpx}
L.~Calibbi, T.~Li, L.~Mukherjee and M.~A. Schmidt, \emph{{Is Dark Matter the
  origin of the $B\to K \nu\bar\nu$ excess at Belle II?}},
  \href{https://arxiv.org/abs/2502.04900}{{\ttfamily 2502.04900}}.

\bibitem{Lee:2025jky}
J.-P. Lee, \emph{{New physics effects in $R(K^{(*)})$, $B_s\to \mu^+\mu^-$, and
  $B^+\to K^+\nu{\bar \nu}$}},
  \href{https://arxiv.org/abs/2502.06370}{{\ttfamily 2502.06370}}.

\bibitem{Athron:2023hmz}
P.~Athron, R.~Martinez and C.~Sierra, \emph{{B meson anomalies and large
  ${B}^{+}\to {K}^{+}\nu \overline{\nu}$ in non-universal U(1)$'$ models}},
  \href{https://doi.org/10.1007/JHEP02(2024)121}{\emph{JHEP} {\bfseries 02}
  (2024) 121}, [\href{https://arxiv.org/abs/2308.13426}{{\ttfamily
  2308.13426}}].

\bibitem{Rosauro-Alcaraz:2024mvx}
S.~Rosauro-Alcaraz and L.~P.~S. Leal, \emph{{Disentangling left and
  right-handed neutrino effects in $B\rightarrow K^{(*)}\nu \nu $}},
  \href{https://doi.org/10.1140/epjc/s10052-024-13104-x}{\emph{Eur. Phys. J. C}
  {\bfseries 84} (2024) 795},
  [\href{https://arxiv.org/abs/2404.17440}{{\ttfamily 2404.17440}}].

\bibitem{Bolton:2025fsq}
P.~D. Bolton, S.~Fajfer, J.~F. Kamenik and M.~Novoa-Brunet, \emph{{Impact of
  new invisible particles on B{\textrightarrow}K(*)Emiss observables}},
  \href{https://doi.org/10.1103/9rrv-ft75}{\emph{Phys. Rev. D} {\bfseries 112}
  (2025) 035010}, [\href{https://arxiv.org/abs/2503.19025}{{\ttfamily
  2503.19025}}].

\bibitem{Aliev:2025hyp}
T.~M. Aliev, A.~Elpe, L.~Selbuz and I.~Turan, \emph{{Explaining Belle data on
  B{\textrightarrow}K(*){\ensuremath{\nu}}{\ensuremath{\nu}}{\textasciimacron}
  decays via dark Z resonances}},
  \href{https://doi.org/10.1103/6j6r-9vsl}{\emph{Phys. Rev. D} {\bfseries 112}
  (2025) 015025}, [\href{https://arxiv.org/abs/2503.22347}{{\ttfamily
  2503.22347}}].

\bibitem{Fuentes-Martin:2020hvc}
J.~Fuentes-Mart{\'\i}n, G.~Isidori, M.~K{\"o}nig and N.~Selimovi{\'c},
  \emph{{Vector Leptoquarks Beyond Tree Level III: Vector-like Fermions and
  Flavor-Changing Transitions}},
  \href{https://doi.org/10.1103/PhysRevD.102.115015}{\emph{Phys. Rev. D}
  {\bfseries 102} (2020) 115015},
  [\href{https://arxiv.org/abs/2009.11296}{{\ttfamily 2009.11296}}].

\bibitem{Dvali:2001gm}
G.~R. Dvali, G.~Gabadadze, M.~Kolanovic and F.~Nitti, \emph{{The Power of brane
  induced gravity}},
  \href{https://doi.org/10.1103/PhysRevD.64.084004}{\emph{Phys. Rev. D}
  {\bfseries 64} (2001) 084004},
  [\href{https://arxiv.org/abs/hep-ph/0102216}{{\ttfamily hep-ph/0102216}}].

\bibitem{Carena:2002me}
M.~Carena, T.~M.~P. Tait and C.~E.~M. Wagner, \emph{{Branes and Orbifolds are
  Opaque}}, {\emph{Acta Phys. Polon. B} {\bfseries 33} (2002) 2355},
  [\href{https://arxiv.org/abs/hep-ph/0207056}{{\ttfamily hep-ph/0207056}}].

\bibitem{delAguila:2003bh}
F.~del Aguila, M.~Perez-Victoria and J.~Santiago, \emph{{Bulk fields with
  general brane kinetic terms}},
  \href{https://doi.org/10.1088/1126-6708/2003/02/051}{\emph{JHEP} {\bfseries
  02} (2003) 051}, [\href{https://arxiv.org/abs/hep-th/0302023}{{\ttfamily
  hep-th/0302023}}].

\bibitem{delAguila:2003kd}
F.~del Aguila, M.~Perez-Victoria and J.~Santiago, \emph{{Some consequences of
  brane kinetic terms for bulk fermions}},  in \emph{{38th Rencontres de
  Moriond on Electroweak Interactions and Unified Theories}}, 5, 2003,
  \href{https://arxiv.org/abs/hep-ph/0305119}{{\ttfamily hep-ph/0305119}}.

\bibitem{delAguila:2003gv}
F.~del Aguila, M.~Perez-Victoria and J.~Santiago, \emph{{Physics of brane
  kinetic terms}}, {\emph{Acta Phys. Polon. B} {\bfseries 34} (2003)
  5511--5522}, [\href{https://arxiv.org/abs/hep-ph/0310353}{{\ttfamily
  hep-ph/0310353}}].

\bibitem{Schwinn:2004xa}
C.~Schwinn, \emph{{Higgsless fermion masses and unitarity}},
  \href{https://doi.org/10.1103/PhysRevD.69.116005}{\emph{Phys. Rev. D}
  {\bfseries 69} (2004) 116005},
  [\href{https://arxiv.org/abs/hep-ph/0402118}{{\ttfamily hep-ph/0402118}}].

\bibitem{Flacke:2008ne}
T.~Flacke, A.~Menon and D.~J. Phalen, \emph{{Non-minimal universal extra
  dimensions}}, \href{https://doi.org/10.1103/PhysRevD.79.056009}{\emph{Phys.
  Rev. D} {\bfseries 79} (2009) 056009},
  [\href{https://arxiv.org/abs/0811.1598}{{\ttfamily 0811.1598}}].

\bibitem{Datta:2012xy}
A.~Datta, U.~K. Dey, A.~Shaw and A.~Raychaudhuri, \emph{{Universal
  Extra-Dimensional Models with Boundary Localized Kinetic Terms: Probing at
  the LHC}}, \href{https://doi.org/10.1103/PhysRevD.87.076002}{\emph{Phys. Rev.
  D} {\bfseries 87} (2013) 076002},
  [\href{https://arxiv.org/abs/1205.4334}{{\ttfamily 1205.4334}}].

\bibitem{Flacke:2013pla}
T.~Flacke, K.~Kong and S.~C. Park, \emph{{Phenomenology of Universal Extra
  Dimensions with Bulk-Masses and Brane-Localized Terms}},
  \href{https://doi.org/10.1007/JHEP05(2013)111}{\emph{JHEP} {\bfseries 05}
  (2013) 111}, [\href{https://arxiv.org/abs/1303.0872}{{\ttfamily 1303.0872}}].

\bibitem{Appelquist:2000nn}
T.~Appelquist, H.-C. Cheng and B.~A. Dobrescu, \emph{{Bounds on universal extra
  dimensions}}, \href{https://doi.org/10.1103/PhysRevD.64.035002}{\emph{Phys.
  Rev. D} {\bfseries 64} (2001) 035002},
  [\href{https://arxiv.org/abs/hep-ph/0012100}{{\ttfamily hep-ph/0012100}}].

\bibitem{Servant:2002hb}
G.~Servant and T.~M.~P. Tait, \emph{{Elastic Scattering and Direct Detection of
  Kaluza-Klein Dark Matter}},
  \href{https://doi.org/10.1088/1367-2630/4/1/399}{\emph{New J. Phys.}
  {\bfseries 4} (2002) 99},
  [\href{https://arxiv.org/abs/hep-ph/0209262}{{\ttfamily hep-ph/0209262}}].

\bibitem{Servant:2002aq}
G.~Servant and T.~M.~P. Tait, \emph{{Is the lightest Kaluza-Klein particle a
  viable dark matter candidate?}},
  \href{https://doi.org/10.1016/S0550-3213(02)01012-X}{\emph{Nucl. Phys. B}
  {\bfseries 650} (2003) 391--419},
  [\href{https://arxiv.org/abs/hep-ph/0206071}{{\ttfamily hep-ph/0206071}}].

\bibitem{Cheng:2002ej}
H.-C. Cheng, J.~L. Feng and K.~T. Matchev, \emph{{Kaluza-Klein dark matter}},
  \href{https://doi.org/10.1103/PhysRevLett.89.211301}{\emph{Phys. Rev. Lett.}
  {\bfseries 89} (2002) 211301},
  [\href{https://arxiv.org/abs/hep-ph/0207125}{{\ttfamily hep-ph/0207125}}].

\bibitem{Majumdar:2002mw}
D.~Majumdar, \emph{{Detection rates for Kaluza-Klein dark matter}},
  \href{https://doi.org/10.1103/PhysRevD.67.095010}{\emph{Phys. Rev. D}
  {\bfseries 67} (2003) 095010},
  [\href{https://arxiv.org/abs/hep-ph/0209277}{{\ttfamily hep-ph/0209277}}].

\bibitem{Burnell:2005hm}
F.~Burnell and G.~D. Kribs, \emph{{The Abundance of Kaluza-Klein dark matter
  with coannihilation}},
  \href{https://doi.org/10.1103/PhysRevD.73.015001}{\emph{Phys. Rev. D}
  {\bfseries 73} (2006) 015001},
  [\href{https://arxiv.org/abs/hep-ph/0509118}{{\ttfamily hep-ph/0509118}}].

\bibitem{Kong:2005hn}
K.~Kong and K.~T. Matchev, \emph{{Precise calculation of the relic density of
  Kaluza-Klein dark matter in universal extra dimensions}},
  \href{https://doi.org/10.1088/1126-6708/2006/01/038}{\emph{JHEP} {\bfseries
  01} (2006) 038}, [\href{https://arxiv.org/abs/hep-ph/0509119}{{\ttfamily
  hep-ph/0509119}}].

\bibitem{Kakizaki:2006dz}
M.~Kakizaki, S.~Matsumoto and M.~Senami, \emph{{Relic abundance of dark matter
  in the minimal universal extra dimension model}},
  \href{https://doi.org/10.1103/PhysRevD.74.023504}{\emph{Phys. Rev. D}
  {\bfseries 74} (2006) 023504},
  [\href{https://arxiv.org/abs/hep-ph/0605280}{{\ttfamily hep-ph/0605280}}].

\bibitem{Belanger:2010yx}
G.~Belanger, M.~Kakizaki and A.~Pukhov, \emph{{Dark matter in UED: The Role of
  the second KK level}},
  \href{https://doi.org/10.1088/1475-7516/2011/02/009}{\emph{JCAP} {\bfseries
  02} (2011) 009}, [\href{https://arxiv.org/abs/1012.2577}{{\ttfamily
  1012.2577}}].

\bibitem{Dienes:1998vh}
K.~R. Dienes, E.~Dudas and T.~Gherghetta, \emph{{Extra space-time dimensions
  and unification}},
  \href{https://doi.org/10.1016/S0370-2693(98)00977-0}{\emph{Phys. Lett. B}
  {\bfseries 436} (1998) 55--65},
  [\href{https://arxiv.org/abs/hep-ph/9803466}{{\ttfamily hep-ph/9803466}}].

\bibitem{Dienes:1998vg}
K.~R. Dienes, E.~Dudas and T.~Gherghetta, \emph{{Grand unification at
  intermediate mass scales through extra dimensions}},
  \href{https://doi.org/10.1016/S0550-3213(98)00669-5}{\emph{Nucl. Phys. B}
  {\bfseries 537} (1999) 47--108},
  [\href{https://arxiv.org/abs/hep-ph/9806292}{{\ttfamily hep-ph/9806292}}].

\bibitem{Bhattacharyya:2006ym}
G.~Bhattacharyya, A.~Datta, S.~K. Majee and A.~Raychaudhuri, \emph{{Power law
  blitzkrieg in universal extra dimension scenarios}},
  \href{https://doi.org/10.1016/j.nuclphysb.2006.10.018}{\emph{Nucl. Phys. B}
  {\bfseries 760} (2007) 117--127},
  [\href{https://arxiv.org/abs/hep-ph/0608208}{{\ttfamily hep-ph/0608208}}].

\bibitem{Hsieh:2006qe}
K.~Hsieh, R.~N. Mohapatra and S.~Nasri, \emph{{Dark matter in universal extra
  dimension models: Kaluza-Klein photon and right-handed neutrino admixture}},
  \href{https://doi.org/10.1103/PhysRevD.74.066004}{\emph{Phys. Rev. D}
  {\bfseries 74} (2006) 066004},
  [\href{https://arxiv.org/abs/hep-ph/0604154}{{\ttfamily hep-ph/0604154}}].

\bibitem{Fujimoto:2014fka}
Y.~Fujimoto, K.~Nishiwaki, M.~Sakamoto and R.~Takahashi, \emph{{Realization of
  lepton masses and mixing angles from point interactions in an extra
  dimension}}, \href{https://doi.org/10.1007/JHEP10(2014)191}{\emph{JHEP}
  {\bfseries 10} (2014) 191},
  [\href{https://arxiv.org/abs/1405.5872}{{\ttfamily 1405.5872}}].

\bibitem{Huang:2012kz}
G.-Y. Huang, K.~Kong and S.~C. Park, \emph{{Bounds on the Fermion-Bulk Masses
  in Models with Universal Extra Dimensions}},
  \href{https://doi.org/10.1007/JHEP06(2012)099}{\emph{JHEP} {\bfseries 06}
  (2012) 099}, [\href{https://arxiv.org/abs/1204.0522}{{\ttfamily 1204.0522}}].

\bibitem{Cheng:2002iz}
H.-C. Cheng, K.~T. Matchev and M.~Schmaltz, \emph{{Radiative corrections to
  Kaluza-Klein masses}},
  \href{https://doi.org/10.1103/PhysRevD.66.036005}{\emph{Phys. Rev. D}
  {\bfseries 66} (2002) 036005},
  [\href{https://arxiv.org/abs/hep-ph/0204342}{{\ttfamily hep-ph/0204342}}].

\bibitem{Flacke:2013nta}
T.~Flacke, K.~Kong and S.~C. Park, \emph{{126 GeV Higgs in Next-to-Minimal
  Universal Extra Dimensions}},
  \href{https://doi.org/10.1016/j.physletb.2013.11.046}{\emph{Phys. Lett. B}
  {\bfseries 728} (2014) 262--267},
  [\href{https://arxiv.org/abs/1309.7077}{{\ttfamily 1309.7077}}].

\bibitem{Bonnevier:2011km}
J.~Bonnevier, H.~Melbeus, A.~Merle and T.~Ohlsson, \emph{{Monoenergetic
  Gamma-Rays from Non-Minimal Kaluza-Klein Dark Matter Annihilations}},
  \href{https://doi.org/10.1103/PhysRevD.85.043524}{\emph{Phys. Rev. D}
  {\bfseries 85} (2012) 043524},
  [\href{https://arxiv.org/abs/1104.1430}{{\ttfamily 1104.1430}}].

\bibitem{Datta:2013nua}
A.~Datta, U.~K. Dey, A.~Raychaudhuri and A.~Shaw, \emph{{Boundary Localized
  Terms in Universal Extra-Dimensional Models through a Dark Matter
  perspective}}, \href{https://doi.org/10.1103/PhysRevD.88.016011}{\emph{Phys.
  Rev. D} {\bfseries 88} (2013) 016011},
  [\href{https://arxiv.org/abs/1305.4507}{{\ttfamily 1305.4507}}].

\bibitem{Dey:2013cqa}
U.~K. Dey and T.~S. Ray, \emph{{Constraining minimal and nonminimal universal
  extra dimension models with Higgs couplings}},
  \href{https://doi.org/10.1103/PhysRevD.88.056016}{\emph{Phys. Rev. D}
  {\bfseries 88} (2013) 056016},
  [\href{https://arxiv.org/abs/1305.1016}{{\ttfamily 1305.1016}}].

\bibitem{Datta:2012tv}
A.~Datta, K.~Nishiwaki and S.~Niyogi, \emph{{Non-minimal Universal Extra
  Dimensions: The Strongly Interacting Sector at the Large Hadron Collider}},
  \href{https://doi.org/10.1007/JHEP11(2012)154}{\emph{JHEP} {\bfseries 11}
  (2012) 154}, [\href{https://arxiv.org/abs/1206.3987}{{\ttfamily 1206.3987}}].

\bibitem{Datta:2013yaa}
A.~Datta, K.~Nishiwaki and S.~Niyogi, \emph{{Non-minimal Universal Extra
  Dimensions with Brane Local Terms: The Top Quark Sector}},
  \href{https://doi.org/10.1007/JHEP01(2014)104}{\emph{JHEP} {\bfseries 01}
  (2014) 104}, [\href{https://arxiv.org/abs/1310.6994}{{\ttfamily 1310.6994}}].

\bibitem{Datta:2013lja}
A.~Datta, A.~Raychaudhuri and A.~Shaw, \emph{{LHC limits on KK-parity
  non-conservation in the strong sector of universal extra-dimension models}},
  \href{https://doi.org/10.1016/j.physletb.2014.01.027}{\emph{Phys. Lett. B}
  {\bfseries 730} (2014) 42--49},
  [\href{https://arxiv.org/abs/1310.2021}{{\ttfamily 1310.2021}}].

\bibitem{Shaw:2014gba}
A.~Shaw, \emph{{KK-parity non-conservation in UED confronts LHC data}},
  \href{https://doi.org/10.1140/epjc/s10052-014-3245-0}{\emph{Eur. Phys. J. C}
  {\bfseries 75} (2015) 33}, [\href{https://arxiv.org/abs/1405.3139}{{\ttfamily
  1405.3139}}].

\bibitem{Shaw:2017whr}
A.~Shaw, \emph{{Status of exclusion limits of the KK-parity non-conserving
  resonance production with updated 13 TeV LHC}},
  \href{https://doi.org/10.5506/APhysPolB.49.1421}{\emph{Acta Phys. Polon. B}
  {\bfseries 49} (2018) 1421},
  [\href{https://arxiv.org/abs/1709.08077}{{\ttfamily 1709.08077}}].

\bibitem{Ganguly:2018pzs}
N.~Ganguly and A.~Datta, \emph{{Exploring non minimal Universal Extra
  Dimensional Model at the LHC}},
  \href{https://doi.org/10.1007/JHEP10(2018)072}{\emph{JHEP} {\bfseries 10}
  (2018) 072}, [\href{https://arxiv.org/abs/1808.08801}{{\ttfamily
  1808.08801}}].

\bibitem{Jha:2014faa}
T.~Jha and A.~Datta, \emph{{$ Z\to b\overline{b} $ in non-minimal Universal
  Extra Dimensional Model}},
  \href{https://doi.org/10.1007/JHEP03(2015)012}{\emph{JHEP} {\bfseries 03}
  (2015) 012}, [\href{https://arxiv.org/abs/1410.5098}{{\ttfamily 1410.5098}}].

\bibitem{Datta:2015aka}
A.~Datta and A.~Shaw, \emph{{Nonminimal universal extra dimensional model
  confronts B$_s\to \mu^+\mu^-$}},
  \href{https://doi.org/10.1103/PhysRevD.93.055048}{\emph{Phys. Rev. D}
  {\bfseries 93} (2016) 055048},
  [\href{https://arxiv.org/abs/1506.08024}{{\ttfamily 1506.08024}}].

\bibitem{Datta:2016flx}
A.~Datta and A.~Shaw, \emph{{Effects of non-minimal Universal Extra Dimension
  on $B\rightarrow X_s\gamma$}},
  \href{https://doi.org/10.1103/PhysRevD.95.015033}{\emph{Phys. Rev. D}
  {\bfseries 95} (2017) 015033},
  [\href{https://arxiv.org/abs/1610.09924}{{\ttfamily 1610.09924}}].

\bibitem{Shaw:2019fin}
A.~Shaw, \emph{{Looking for $B\rightarrow X_s \ell^+\ell^-$ in a nonminimal
  universal extra dimensional model}},
  \href{https://doi.org/10.1103/PhysRevD.99.115030}{\emph{Phys. Rev. D}
  {\bfseries 99} (2019) 115030},
  [\href{https://arxiv.org/abs/1903.10302}{{\ttfamily 1903.10302}}].

\bibitem{Biswas:2017vhc}
A.~Biswas, A.~Shaw and S.~K. Patra, \emph{{$\mathcal{R}(D^{(*)})$ anomalies in
  light of a nonminimal universal extra dimension}},
  \href{https://doi.org/10.1103/PhysRevD.97.035019}{\emph{Phys. Rev. D}
  {\bfseries 97} (2018) 035019},
  [\href{https://arxiv.org/abs/1708.08938}{{\ttfamily 1708.08938}}].

\bibitem{Dasgupta:2018nzt}
S.~Dasgupta, U.~K. Dey, T.~Jha and T.~S. Ray, \emph{{Status of a flavor-maximal
  nonminimal universal extra dimension model}},
  \href{https://doi.org/10.1103/PhysRevD.98.055006}{\emph{Phys. Rev. D}
  {\bfseries 98} (2018) 055006},
  [\href{https://arxiv.org/abs/1801.09722}{{\ttfamily 1801.09722}}].

\bibitem{Lee:2019phf}
J.-P. Lee, \emph{{$B$ anomalies in the nonminimal universal extra dimension
  model}}, \href{https://doi.org/10.1103/PhysRevD.100.075005}{\emph{Phys. Rev.
  D} {\bfseries 100} (2019) 075005},
  [\href{https://arxiv.org/abs/1906.07345}{{\ttfamily 1906.07345}}].

\bibitem{Dey:2016cve}
U.~K. Dey and T.~Jha, \emph{{Rare top decays in minimal and nonminimal
  universal extra dimension models}},
  \href{https://doi.org/10.1103/PhysRevD.94.056011}{\emph{Phys. Rev. D}
  {\bfseries 94} (2016) 056011},
  [\href{https://arxiv.org/abs/1602.03286}{{\ttfamily 1602.03286}}].

\bibitem{Chiang:2018oyd}
C.-W. Chiang, U.~K. Dey and T.~Jha, \emph{{$t \rightarrow cg$ and $t
  \rightarrow cZ$ in universal extra-dimensional models}},
  \href{https://doi.org/10.1140/epjp/i2019-12607-1}{\emph{Eur. Phys. J. Plus}
  {\bfseries 134} (2019) 210},
  [\href{https://arxiv.org/abs/1807.01481}{{\ttfamily 1807.01481}}].

\bibitem{Shaw:2020fxf}
A.~Shaw, \emph{{The impact of nonminimal Universal Extra Dimensional model on
  $\Delta B=2$ transitions}},
  \href{https://doi.org/10.1140/epjc/s10052-021-08937-9}{\emph{Eur. Phys. J. C}
  {\bfseries 81} (2021) 137},
  [\href{https://arxiv.org/abs/2006.12164}{{\ttfamily 2006.12164}}].

\bibitem{Jha:2016sre}
T.~Jha, \emph{{Unitarity Constraints on non-minimal Universal Extra Dimensional
  Model}}, \href{https://doi.org/10.1088/1361-6471/aade80}{\emph{J. Phys. G}
  {\bfseries 45} (2018) 115002},
  [\href{https://arxiv.org/abs/1604.02481}{{\ttfamily 1604.02481}}].

\bibitem{Datta:2014sha}
A.~Datta and A.~Shaw, \emph{{A note on gauge-fixing in the electroweak sector
  of non-minimal UED}},
  \href{https://doi.org/10.1142/S0217732316501819}{\emph{Mod. Phys. Lett. A}
  {\bfseries 31} (2016) 1650181},
  [\href{https://arxiv.org/abs/1408.0635}{{\ttfamily 1408.0635}}].

\bibitem{Colangelo:1996ay}
P.~Colangelo, F.~De~Fazio, P.~Santorelli and E.~Scrimieri, \emph{{Rare $B \to
  K^{(*)}$ neutrino anti-neutrino decays at $B$ factories}},
  \href{https://doi.org/10.1016/S0370-2693(97)00130-5}{\emph{Phys. Lett. B}
  {\bfseries 395} (1997) 339--344},
  [\href{https://arxiv.org/abs/hep-ph/9610297}{{\ttfamily hep-ph/9610297}}].

\bibitem{Altmannshofer:2009ma}
W.~Altmannshofer, A.~J. Buras, D.~M. Straub and M.~Wick, \emph{{New strategies
  for New Physics search in $B \to K^{*} \nu \bar{\nu}$, $B \to K \nu
  \bar{\nu}$ and $B \to X_{s} \nu \bar{\nu}$ decays}},
  \href{https://doi.org/10.1088/1126-6708/2009/04/022}{\emph{JHEP} {\bfseries
  04} (2009) 022}, [\href{https://arxiv.org/abs/0902.0160}{{\ttfamily
  0902.0160}}].

\bibitem{Buras:2014fpa}
A.~J. Buras, J.~Girrbach-Noe, C.~Niehoff and D.~M. Straub, \emph{{$ B\to
  {K}^{\left(\ast \right)}\nu \overline{\nu} $ decays in the Standard Model and
  beyond}}, \href{https://doi.org/10.1007/JHEP02(2015)184}{\emph{JHEP}
  {\bfseries 02} (2015) 184},
  [\href{https://arxiv.org/abs/1409.4557}{{\ttfamily 1409.4557}}].

\bibitem{Buras:2002ej}
A.~J. Buras, M.~Spranger and A.~Weiler, \emph{{The Impact of universal extra
  dimensions on the unitarity triangle and rare K and B decays}},
  \href{https://doi.org/10.1016/S0550-3213(03)00250-5}{\emph{Nucl. Phys. B}
  {\bfseries 660} (2003) 225--268},
  [\href{https://arxiv.org/abs/hep-ph/0212143}{{\ttfamily hep-ph/0212143}}].

\bibitem{Buchalla:1993bv}
G.~Buchalla and A.~J. Buras, \emph{{QCD corrections to rare K and B decays for
  arbitrary top quark mass}},
  \href{https://doi.org/10.1016/0550-3213(93)90405-E}{\emph{Nucl. Phys. B}
  {\bfseries 400} (1993) 225--239}.

\bibitem{Buchalla:1998ba}
G.~Buchalla and A.~J. Buras, \emph{{The rare decays $K\to \pi \nu\bar\nu$, $B
  \to X \nu\bar\nu$ and $B \to l^+ l^-$: An Update}},
  \href{https://doi.org/10.1016/S0550-3213(99)00149-2}{\emph{Nucl. Phys. B}
  {\bfseries 548} (1999) 309--327},
  [\href{https://arxiv.org/abs/hep-ph/9901288}{{\ttfamily hep-ph/9901288}}].

\bibitem{Misiak:1999yg}
M.~Misiak and J.~Urban, \emph{{QCD corrections to FCNC decays mediated by Z
  penguins and W boxes}},
  \href{https://doi.org/10.1016/S0370-2693(99)00150-1}{\emph{Phys. Lett. B}
  {\bfseries 451} (1999) 161--169},
  [\href{https://arxiv.org/abs/hep-ph/9901278}{{\ttfamily hep-ph/9901278}}].

\bibitem{Brod:2010hi}
J.~Brod, M.~Gorbahn and E.~Stamou, \emph{{Two-Loop Electroweak Corrections for
  the $K \to \pi \nu \bar{\nu}$ Decays}},
  \href{https://doi.org/10.1103/PhysRevD.83.034030}{\emph{Phys. Rev. D}
  {\bfseries 83} (2011) 034030},
  [\href{https://arxiv.org/abs/1009.0947}{{\ttfamily 1009.0947}}].

\bibitem{ParticleDataGroup:2024cfk}
{\scshape Particle Data Group} collaboration, S.~Navas et~al., \emph{{Review of
  particle physics}},
  \href{https://doi.org/10.1103/PhysRevD.110.030001}{\emph{Phys. Rev. D}
  {\bfseries 110} (2024) 030001}.

\bibitem{Buchalla:1995vs}
G.~Buchalla, A.~J. Buras and M.~E. Lautenbacher, \emph{{Weak Decays beyond
  Leading Logarithms}},
  \href{https://doi.org/10.1103/RevModPhys.68.1125}{\emph{Rev. Mod. Phys.}
  {\bfseries 68} (1996) 1125--1144},
  [\href{https://arxiv.org/abs/hep-ph/9512380}{{\ttfamily hep-ph/9512380}}].

\bibitem{Belle:2017oht}
{\scshape Belle} collaboration, J.~Grygier et~al., \emph{{Search for
  $\boldsymbol{B\to h\nu\bar{\nu}}$ decays with semileptonic tagging at
  Belle}}, \href{https://doi.org/10.1103/PhysRevD.96.091101}{\emph{Phys. Rev.
  D} {\bfseries 96} (2017) 091101},
  [\href{https://arxiv.org/abs/1702.03224}{{\ttfamily 1702.03224}}].

\bibitem{Datta:2012db}
A.~Datta and S.~Raychaudhuri, \emph{{Vacuum Stability Constraints and LHC
  Searches for a Model with a Universal Extra Dimension}},
  \href{https://doi.org/10.1103/PhysRevD.87.035018}{\emph{Phys. Rev. D}
  {\bfseries 87} (2013) 035018},
  [\href{https://arxiv.org/abs/1207.0476}{{\ttfamily 1207.0476}}].

\bibitem{Dey:2004gb}
P.~Dey and G.~Bhattacharyya, \emph{{A Comparison of ultraviolet sensitivities
  in universal, nonuniversal, and split extra dimensional models}},
  \href{https://doi.org/10.1103/PhysRevD.70.116012}{\emph{Phys. Rev. D}
  {\bfseries 70} (2004) 116012},
  [\href{https://arxiv.org/abs/hep-ph/0407314}{{\ttfamily hep-ph/0407314}}].

\bibitem{Straub:2018kue}
D.~M. Straub, \emph{{flavio: a Python package for flavour and precision
  phenomenology in the Standard Model and beyond}},
  \href{https://arxiv.org/abs/1810.08132}{{\ttfamily 1810.08132}}.

\bibitem{HFLAV:2022esi}
{\scshape HFLAV} collaboration, Y.~S. Amhis et~al., \emph{{Averages of
  b-hadron, c-hadron, and {\ensuremath{\tau}}-lepton properties as of 2021}},
  \href{https://doi.org/10.1103/PhysRevD.107.052008}{\emph{Phys. Rev. D}
  {\bfseries 107} (2023) 052008},
  [\href{https://arxiv.org/abs/2206.07501}{{\ttfamily 2206.07501}}].

\bibitem{Lees:2013nxa}
{\scshape BaBar} collaboration, J.~P. Lees et~al., \emph{{Measurement of the $B
  \to X_s l^+l^-$ branching fraction and search for direct CP violation from a
  sum of exclusive final states}},
  \href{https://doi.org/10.1103/PhysRevLett.112.211802}{\emph{Phys. Rev. Lett.}
  {\bfseries 112} (2014) 211802},
  [\href{https://arxiv.org/abs/1312.5364}{{\ttfamily 1312.5364}}].

\bibitem{Deutschmann:2017bth}
N.~Deutschmann, T.~Flacke and J.~S. Kim, \emph{{Current LHC Constraints on
  Minimal Universal Extra Dimensions}},
  \href{https://doi.org/10.1016/j.physletb.2017.06.004}{\emph{Phys. Lett. B}
  {\bfseries 771} (2017) 515--520},
  [\href{https://arxiv.org/abs/1702.00410}{{\ttfamily 1702.00410}}].

\bibitem{Ball:2004ye}
P.~Ball and R.~Zwicky, \emph{{New results on $B \to \pi, K, \eta$ decay
  formfactors from light-cone sum rules}},
  \href{https://doi.org/10.1103/PhysRevD.71.014015}{\emph{Phys. Rev. D}
  {\bfseries 71} (2005) 014015},
  [\href{https://arxiv.org/abs/hep-ph/0406232}{{\ttfamily hep-ph/0406232}}].

\bibitem{Bartsch:2009qp}
M.~Bartsch, M.~Beylich, G.~Buchalla and D.~N. Gao, \emph{{Precision Flavour
  Physics with $B \to K \nu \bar\nu$ and $B \to K l^+ l^-$}},
  \href{https://doi.org/10.1088/1126-6708/2009/11/011}{\emph{JHEP} {\bfseries
  11} (2009) 011}, [\href{https://arxiv.org/abs/0909.1512}{{\ttfamily
  0909.1512}}].

\bibitem{FlavourLatticeAveragingGroupFLAG:2021npn}
{\scshape Flavour Lattice Averaging Group (FLAG)} collaboration, Y.~Aoki
  et~al., \emph{{FLAG Review 2021}},
  \href{https://doi.org/10.1140/epjc/s10052-022-10536-1}{\emph{Eur. Phys. J. C}
  {\bfseries 82} (2022) 869},
  [\href{https://arxiv.org/abs/2111.09849}{{\ttfamily 2111.09849}}].

\bibitem{Altmannshofer:2008dz}
W.~Altmannshofer, P.~Ball, A.~Bharucha, A.~J. Buras, D.~M. Straub and M.~Wick,
  \emph{{Symmetries and Asymmetries of $B \to K^{*} \mu^{+} \mu^{-}$ Decays in
  the Standard Model and Beyond}},
  \href{https://doi.org/10.1088/1126-6708/2009/01/019}{\emph{JHEP} {\bfseries
  01} (2009) 019}, [\href{https://arxiv.org/abs/0811.1214}{{\ttfamily
  0811.1214}}].

\bibitem{Bharucha:2015bzk}
A.~Bharucha, D.~M. Straub and R.~Zwicky, \emph{{$B\to V\ell^+\ell^-$ in the
  Standard Model from light-cone sum rules}},
  \href{https://doi.org/10.1007/JHEP08(2016)098}{\emph{JHEP} {\bfseries 08}
  (2016) 098}, [\href{https://arxiv.org/abs/1503.05534}{{\ttfamily
  1503.05534}}].

\end{thebibliography}
\end{document}